\documentclass{aa}  

\usepackage{graphicx}
\usepackage{txfonts}
\usepackage[colorlinks,linkcolor=blue,citecolor=blue]{hyperref}

\usepackage{orcidlink}
\usepackage{mathrsfs}
\usepackage[cal=cm]{mathalpha}
\usepackage{natbib}
\bibpunct{(}{)}{;}{a}{}{,}

\begin{document} 

   \title{Gravity versus astrophysics in black hole images and photon rings: Equatorial emissions and spherically symmetric space-times}
   \titlerunning{Degeneracy between gravity and astrophysics in photon rings}

   \author{I. Urso
          \inst{1,2}\orcidlink{0009-0003-2105-522X}
          \and
          F. H. Vincent\inst{1}\orcidlink{0000-0002-9214-0830}
          \and M. Wielgus\inst{3}\orcidlink{0000-0002-8635-4242}
          \and T. Paumard\inst{1}\orcidlink{0000-0003-0655-0452}
          \and G. Perrin\inst{1}\orcidlink{0000-0003-0680-0167}
          }

   \institute{LIRA, Observatoire de Paris, Université PSL, CNRS, Sorbonne Université, Université Paris Cité, CY Cergy Paris Université, 5 Place Jules Janssen, 92190 Meudon, France\\
   \email{irene.urso@obspm.fr}
    \and
        Laboratoire de Physique de l’École normale supérieure, ENS, F-75005 Paris, France
    \and Instituto de Astrof\'{i}sica de Andaluc\'{i}a-CSIC, Glorieta de la Astronom\'{i}a s/n, E-18008 Granada, Spain
        }

   \date{Received March 31, 2025; accepted June 28, 2025}

  \abstract 
   {The Event Horizon Telescope (EHT) collaboration released in 2019 the first horizon-scale images of a black hole accretion flow, opening a novel route for plasma physics comprehension and gravitational tests. Although the present unresolved images deeply depend on the astrophysical properties of the accreted matter, general relativity predicts that they contain highly lensed observables, the so-called photon rings, embodying the effects of strong-field gravity.}
   {Focusing on the particular case of the supermassive black hole M87* and adopting a geometrically thin equatorial disc as a phenomenological configuration for the accreting matter, our goal is to study the degeneracy of space-time curvature and of physically motivated emission processes on plane-of-sky EHT-like images observed at 230 and 345 GHz.}
   {In a parametric framework, we simulated adaptively ray-traced images using the code GYOTO in various spherically symmetric space-time geometries for a comprehensive class of disc velocities and a library of realistic synchrotron emission profiles. We then extracted the width and the peak position of 1D intensity cross sections on the direct image and the first photon ring.}
   {We show that among the investigated quantities, the most appropriate observables to probe the geometry are the peak positions of the first photon ring. Small geometric deviations can be unequivocally detected regardless of the motion of the disc, ranging from Keplerian rotation to radial infall, if the black hole mass-to-distance estimate is accurate up to around 2\%, with the current uncertainty of 11\% being just sufficient to access extreme deviations.}
   {The equatorial set-up of this paper, which is favoured by present EHT observations of M87*, is adapted to modelling future measurements at higher observing frequencies, where absorption effects are negligible, and with higher resolution, indispensable to resolving the photon rings. Additional work is needed to investigate if our conclusions hold for more realistic disc configurations.}


   \maketitle

\section{Introduction}

The existence of black holes is a direct consequence of general relativity, the standard theory of gravity, and their first solution was found by Karl Schwarzschild \citep{schwarzschild} a few weeks after the publication of the general relativity foundations.
A cardinal generalisation to rotating black holes, published by Roy Kerr \citep{kerr}, represents the only\footnote{This holds in the astrophysically relevant case of null electric charge.} stationary solution satisfying the hypotheses of the no-hair theorem \citep{israel,carter1971axisymmetric,hawking1972black,robinson1975uniqueness}, and a Kerr black hole is entirely described by its mass and angular momentum.\\
\indent The definition of a black hole relies on the presence of an immaterial boundary, the event horizon, that can be crossed only in one direction, towards the black hole. Notably, not even light can escape it. According to the cosmic censorship conjecture \citep{penrose}, for a Kerr or a Schwarzschild black hole, the event horizon also hides observationally inaccessible incomplete null geodesics that may be associated with a curvature singularity. On top of that, the interior of the Kerr space-time contains closed time-like curves that correspond to a causality violation \citep{carter1968global}. These predictions are controversial, and they might mark the limit of the applicability domain of general relativity. It is then justified to continue testing this centennial theory, which has already been thoroughly corroborated for weak gravitational fields, in the extreme conditions of the strong-field regime begotten by black holes.\\
\indent Since a black hole is not a luminous source itself, its presence is inferred from the influence it has on its environment, which has been revealed by recent revolutionary astronomical observations. The first detection of a gravitational wave signal from the merger of two black holes was announced in 2016 by the LIGO-Virgo collaboration \citep{LIGOVirgo}. Two other great achievements, both by GRAVITY in 2018, were the astrometric measurements, close to the supermassive black hole Sagittarius A*, of the orbital motion of the S stars \citep{GRAVITY} and of the high states, or flares, of variable near-infrared emission \citep{Flares}. Finally, the first breakthrough images of black hole accretion flows probing the strong-field gravity region just above the event horizon were released by the Event Horizon Telescope (EHT) in 2019 for M87* \citep{EHTM87}.\\
\indent These images, although compatible with the GR-predicted observational appearance of a Kerr black hole, do not exclude alternative space-times \citep{EHTVI,vincent2021geometric,gralla2021M87}. However, consistency tests of the Kerr-hypothesis will become more stringent with the upgrades of the Next Generation EHT \citep[ngEHT,][]{ngEHT,doeleman2023reference,ayzenberg2023fundamental} and, particularly, with future space-based very-long-baseline interferometers such as the proposed mission Black Hole Explorer \citep[BHEX,][]{BHEX,hudson2023orbital}, which will achieve sufficient angular resolution to access the most gravitationally affected features of the image, namely the photon rings.\\
\indent It has already been widely shown that the visual aspect of the accretion flow depends both on the space-time geometry and the properties of the emission and that the observational appearance of M87*, under the limited instrumental resolution of the EHT, can be mimicked by non-Kerr objects \citep[for instance,][]{vincent2021geometric}. In this paper we do not attempt to model the limitations of any particular existing or future observing instrument, assuming that a perfect image-domain reconstruction can be obtained. Thus, we focus on the degeneracy between gravity and astrophysics in the images that is fundamental and theoretical in nature. Hence, it is crucial to identify observable components of the images that allow these effects to be disentangled. In this work we focus on photon rings, that is, lensed secondary images of the accretion disc superposed on top of the primary image and formed by photons executing at least a half loop around the black hole. In contrast to the primary image and the central depression in brightness, known as the `observable shadow', which are strongly dependent on the astrophysical configuration  \citep{EHTV,bauer2022spherical,kocherlakota2022distinguishing}, photon rings are predominantly impacted by the space-time geometry \citep{desire2025multifrequency}.\\
\indent Photon rings are labelled by the number of half turns, $n$, of the photons around the central body. It has been shown that they follow a specific functional form for the Kerr metric not only at high $n$ \citep{gralla2020shape} but also at $n=1$ \citep{cardenas2023prediction}. Moreover, as shown in the latter paper for $n=1$ and elsewhere for $n\geq2$ \citep{paugnat2022photon}, the corresponding interferometric signature is robust under change of the astrophysical configuration for a Kerr black hole. A limitation present in these works that we would like to overcome here is the adoption of a three-parameter subset of Johnson’s Standard-Unbounded distribution \citep{gralla2020shape} without physical parameters describing the emission process.\\
\indent Properties of the photon rings have been used not only to carry out the consistency tests mentioned in the previous paragraph but also to investigate discriminatory tests of general relativity, at least for specific classes of metrics and heuristic emission processes. For example, \citet{bauer2022spherical} and \citet{kocherlakota2022distinguishing} have both investigated the properties of the intensity collected on EHT-like images of spherically symmetric static compact objects surrounded by spherically infalling emitting matter. In this framework, thin disc models have also been studied, as in \citet{eichhorn2023universal}. Several other authors have suggested the use of Lyapunov exponents to differentiate between the space-time metrics of compact objects that admit separable geodesic equations \citep{wielgus2021photon,staelens2023black,da2023photon,kocherlakota2024prospects}. However, since that theoretical feature of photon rings is not directly detectable, one should invoke, for instance, their link with potentially measurable brightness autocorrelations
\citep{hadar2021photon,chesler2021light}.\\
\indent Finally, a promising perspective for the study of photon rings consists in the analysis of their polarised signal
\citep{himwich2020universal}, which could lead to the detection of the first photon ring before reaching the baselines
needed to resolve its interferometric imprint \citep{palumbo2023demonstrating}.\\
\indent The purpose of this paper is to propose a discriminatory test of general relativity focusing on a parametrised spherically symmetric space-time model of M87*, as in \citet{bauer2022spherical,kocherlakota2022distinguishing}, and to address the image degeneracy between the nature of the compact object and a large library of physically motivated synchrotron emission profiles for an equatorial disc. We tested Keplerian, radially infalling, and mixed velocity fields, with a special interest in the first secondary image of the flow, the $n=1$ photon ring. The main restrictions of this preliminary work are the adoptions of a non-rotating black hole and of a geometrically thin accretion disc. Nonetheless, the latter hypothesis may be justified very close to the event horizon both in the framework of classic thick disc accretion models in the plunging region, below the cusp \citep{abramowicz1978relativistic}, as well as through more recent arguments based on general relativistic magnetohydrodynamic (GRMHD) simulations \citep{chael2021observing}. Also, we only simulated instantaneous time-averaged images, and we did not explore the associated interferometric visibility as measured by the EHT. These constrictive assumptions can be relaxed in subsequent work by making use of already existing examples for an axisymmetric parametrised space-time \citep{konoplya2016general}, models for thick accretion flows \citep{vincent2022images}, and astrophysical fluctuations \citep{lee2021disks}.\\
\indent This paper is organised as follows. First, Section \hyperref[sec:img]{2} provides
a review of the definitions of the image components. Then, in Sections \hyperref[sec:metric]{3} and \hyperref[sec:disc]{4} we
detail our models for the compact object and the accretion disc. Next, we describe our simulated images in Section \hyperref[sec:simu]{5}, and we analyse some of their distinctive features in Section \hyperref[sec:analysis]{6}. Finally, we comment on the gravity-astrophysics degeneracy in Section \hyperref[sec:degeneracy]{7} and offer a concluding discussion in Section \hyperref[sec:end]{8}.

\section{Schwarzschild image features and observables}
\label{sec:img}

\begin{figure*}[ht!]
   \resizebox{\hsize}{!}
            {\includegraphics{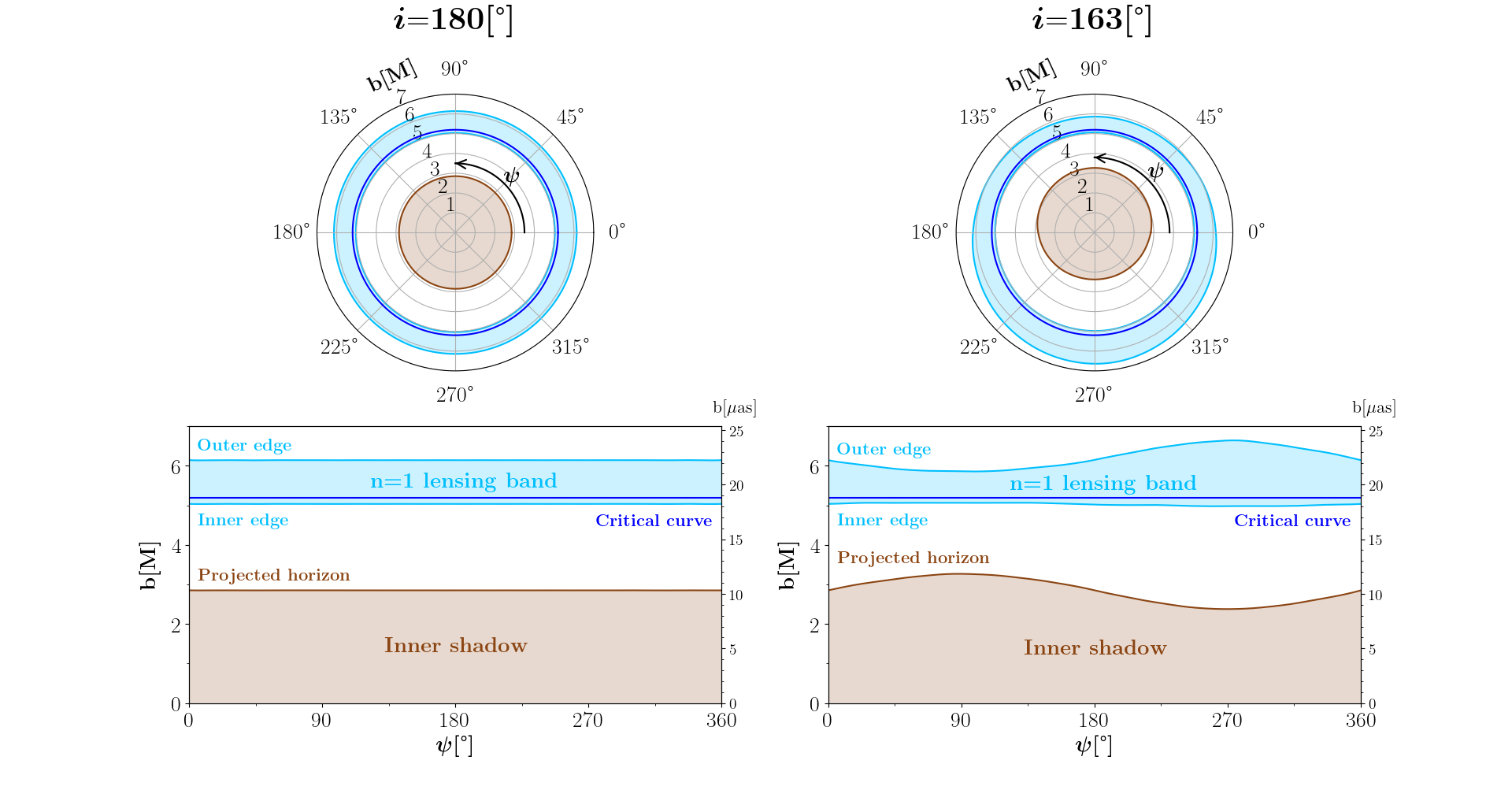}}
    \caption{Critical curve (dark blue), $n=1$ lensing band (light blue), and inner shadow (brown) of a Schwarzschild black hole on the plane of sky (upper panels) seen at different inclinations and their impact parameters as a function of the polar angle on the screen (lower panels).}
\label{fig:theoretical_features}
\end{figure*}

In this paper, we consider static and spherically symmetric parametrised black hole metrics, that is, deformations of the vacuum Schwarzschild solution of general relativity. This section reviews the properties of null geodesics in this standard space-time. However, the notions explored here can be and have been analysed in the spinning case (see \citet{lupsasca2024beginner} for a general introductory review on Kerr black hole imaging). 
Here, all the expressions are written in geometrised units ($G=c=1$), so $M$, the mass of the black hole, has the dimension of a length. We recall that the radial position of a Schwarzschild event horizon, in Schwarzschild-Droste coordinates $(t,r,\theta,\varphi)$, is $r_\mathrm{H}=2M$.

\subsection{Null geodesics and photon shell}

While the equations of geodesic motion in the Schwarzschild\footnote{Their study in the Kerr  metric was initiated by Carter \citep{carter1968global} and it has recently been revisited and completed \citep{gralla2020null}.} space-time were first derived by Schwarzschild \citep{schwarzschild} and Droste \citep{droste1917field}, the explicit examination of null geodesics governing the trajectory of massless particles like photons is, as far as we know, due to Flamm \citep{flamm1916beitrage}.\\ 
\indent The four null geodesic equations can be obtained by means of the expressions translating the symmetries of the space-time.
The two Killing vectors, $\boldsymbol{\partial}_t$ and $\boldsymbol{\partial}_\varphi$, associated with the stationarity and axisymmetry of the Schwarzschild space-time result in conserved energy, $E$, and conserved angular momentum, $L$:
\[ E=-\boldsymbol{\partial}_t\,\cdot\boldsymbol{p}=-p_t\,, \quad \quad L=\boldsymbol{\partial}_\varphi\,\cdot\boldsymbol{p}=p_\varphi\,, \quad \quad \lambda \coloneqq \frac{L}{E}\:,
\tag{1} \label{eq:1}\]
where $p^\mu=\mathrm{d}x^\mu/\mathrm{d}s$ are the components of the 4-momentum of the photon in Schwarzschild-Droste coordinates, with an affine parameter, $s$, that increases along the geodesic towards the future. The energy-rescaled angular momentum, $\lambda$, or equivalently the impact parameter, $b\coloneqq|\lambda|$, is properly defined for null geodesics reaching an observer at asymptotic infinity, the ones of interest here, for which $E>0$, and it entirely characterises them \citep{wald1984general}.
Also, preserving the spherical symmetry confines the photon on a time-like-planar hypersurface and its mass, $m=0$, represents the fourth integral of motion.\\
\indent Excluding worldlines that terminate in the black hole region, there are no stable bound orbits for photons, since $r(s)$ is either a monotonic function or has a single radial turning point. Besides, photons orbit at fixed Schwarzschild-Droste radius, $\tilde{r}$, if and only if, in Schwarzschild-Droste coordinates \citep{hilbert1917grundlagen}
\begin{align*}
	\tilde{r}=3M\quad\text{and} \quad\tilde{b}=3\sqrt{3}M\:.
	\tag{2} \label{eq:2}
\end{align*}
As mentioned above, these circular orbits are unstable: at the slightest perturbation, the photon initially at $\tilde{r}$ either falls inside the event horizon or it escapes to infinity. 
The region of the space-time spanned by all the circular photon orbits is called the photon sphere.\footnote{For Kerr black holes we talk of spherical orbits and photon shell \citep{teo2003spherical}.} A photon winds more and more times, clockwise or anticlockwise according to the sign of $\lambda$, as it approaches the photon sphere, where geodesics loop indefinitely.

\subsection{Critical curve and lensing bands}

The critical curve is the theoretical image-plane closed curve depicted by the impact points of null trajectories possessing the same constant of motion, $\tilde{\lambda}$, as the bound photon orbit that they asymptotically approach. Simplistically, the critical curve, also referred to as `apparent boundary' \citep{bardeen1973timelike}, is the projection of the photon sphere on the plane of sky and it delimits two regions according to the impact parameter, $b$, of the geodesics.
Reversing time's arrow, one can imagine that the photon is launched from the screen of the observer, at $r_\mathrm{em}>\tilde{r}$, towards the black hole: inside the critical curve, that is, in the so-called `black hole shadow' defined by $b<\tilde{b}$, null geodesics cross the horizon, while outside of it, when $b>\tilde{b}$, photon trajectories are not trapped. We stress that `black hole shadow' is an unfortunate naming convention because it does not generally match with the effectively darker region on the image, as it is underlined in Paragraph \hyperref[sec:shadow]{2.4}. For a Schwarzschild black hole, the critical curve is a circle
of radius equal to the critical impact parameter, $\tilde{b}=3\sqrt{3}M\simeq 5.196152M$;\footnote{In the Kerr case, its angular coordinates on the screen of a distant observer were first derived by Cunningham and Bardeen \citep{cunningham1973optical} and have been re-examined in present-day papers \citep{teo2003spherical,johnson2020universal}. Its cardioid shape, of similar size as the Schwarzschild one, depends on spin and inclination \citep{falcke1999viewing}.} we represent the critical curve as a dark blue line in Figure \hyperref[fig:theoretical_features]{1}.\\
\indent Lensing bands are regions on the observer's screen that comprise all the impact points reached by photons with constants of motions close to those of a bound trajectory, that is, with $\lambda\approx\tilde{\lambda}$, and that follow nearly bound null orbits \citep{gralla2019black}: these geodesics pass close to the photon shell, without being trapped in such a way that they can perform several turns around the black hole before leaving to infinity. More precisely, given an equatorial plane passing through the black hole, we say that a light ray belongs to the $n$-th lensing band, with $n\geq1$, if and only if it intersects the given equatorial plane exactly $n+1$ times on its way to the observer \citep{chael2021observing,paugnat2022photon}. If the plane is parallel to the screen, this is equivalent to the statement that the photons of $n$-th lensing band execute $n$ half turns around the black hole before reaching the observer. An alternative definition of lensing bands is presented in Appendix \hyperref[app:lensing_bands]{A}.\\
\indent Lensing bands are perfectly circular for a black hole seen face-on, whereas they are highly deformed for increasing inclinations, $i$, of the distant observer,\footnote{In a Kerr space-time, they also weakly depend on the spin \citep{gralla2020lensing}.} as shown by the light blue bands of Figure \hyperref[fig:theoretical_features]{1}. In particular, lensing bands of an inclined screen possess a reflection symmetry,\footnote{This is only true for spherically symmetric space-times \citep{gralla2020lensing}.} they correspond to those of the non-inclined screen along the line perpendicular to the axis of reflection and their width varies according to the polar angle on the screen (see Appendix \hyperref[app:lensing_bands]{A} for a proof of these statements). Also, the inner edge of the $n$-th lensing band is traced by means of the impact parameters of light rays of order $n$ that asymptotically approach the event horizon, $r\to r_\mathrm{H}$, while its outer edge corresponds to the image of a circle of radius $r\to\infty$ on the plane transported by light rays of order $n$ (see Figure \hyperref[fig:Deflection]{A.3} in Appendix \hyperref[app:lensing_bands]{A}). We highlight that in our plots, the outer edge of the lensing band is prescribed by large, but not infinite, values of $r$ according to numerical stop conditions on the disc extent.\\
\indent The critical curve, as well as the lensing bands defined with respect to a given equatorial plane\footnote{\citet{vincent2022images} and \citet{kocherlakota2024hotspots} offer a discussion on image orders when the crossed shape possesses some thickness.} depend exclusively on the geometry of space-time and are insensitive to the accretion disc properties. Nonetheless, because of the asymptotic character of their precise definitions, they are not the observables that could be directly measured from images of black holes.

\subsection{Apparent shape of circular rings}
\label{par:rings}
\begin{figure*}[ht!]
   \resizebox{\hsize}{!}
            {\includegraphics{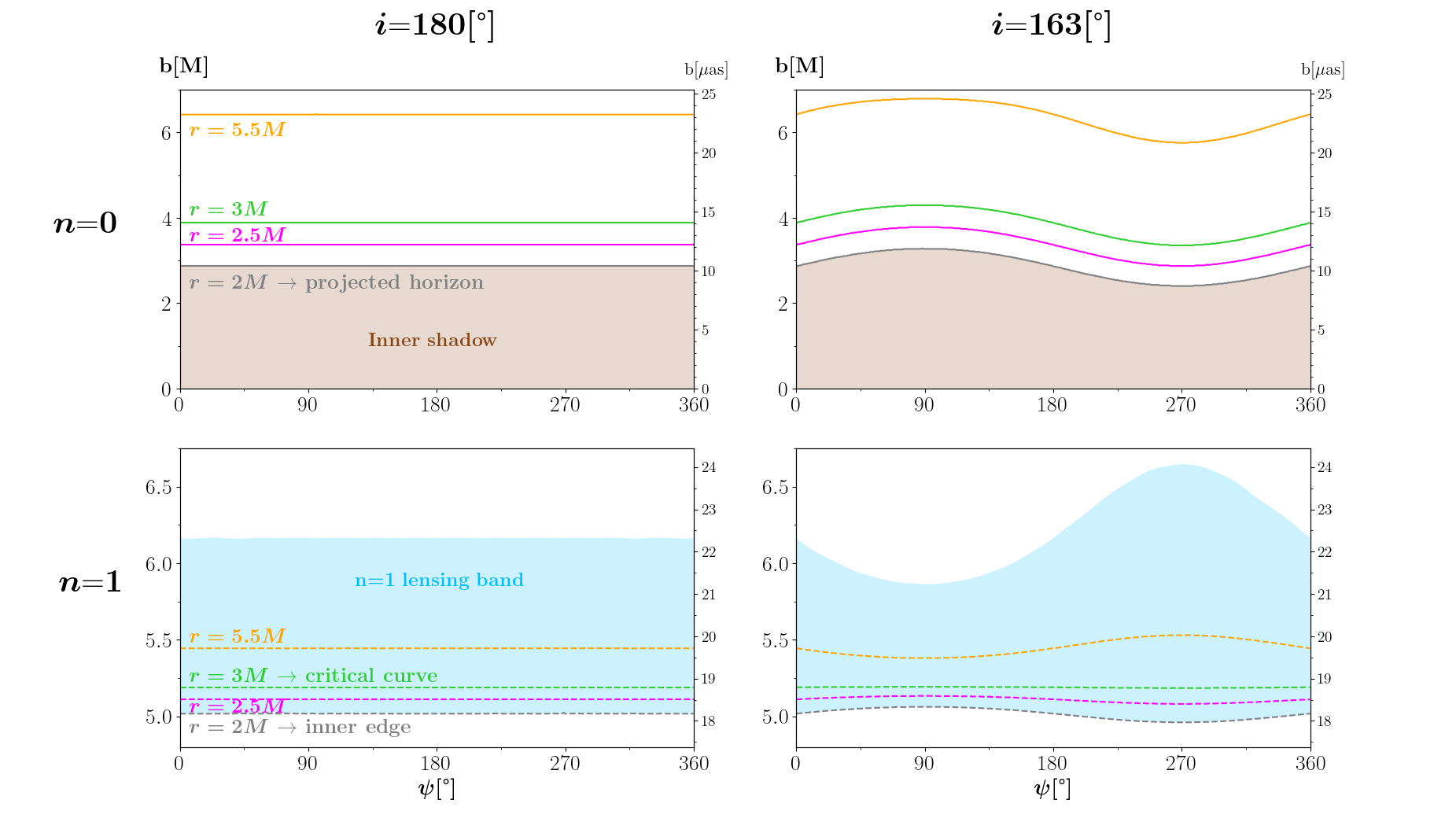}}
    \caption{Direct (solid lines in the upper panels) and first-lensed (dashed lines in the lower panels) apparent positions of rays emitted from isoradial distances, $r$, to the black hole at different inclinations: $r=2M$ (grey), $2.5M$ (magenta), $3M$ (green), and $5.5M$ (yellow). The inner shadow (brown) and the first lensing band (light blue) are represented as well.}
\label{fig:apparent_radii}
\end{figure*}

We analysed the projections of equatorial rings of constant radii, $r$. At this point we were only interested in mapping between a given equatorial plane and the observer’s screen, connected with null geodesics. Hence, we did not need to specify the emission and opacity properties nor the redshift effects. Furthermore, we restrained our attention to light rays of orders $n=0,1$, but the following enquiries can be generalised to a higher order. Discussions presented herein are illustrated in Figure \hyperref[fig:apparent_radii]{2}.\\
\indent If the equatorial plane is observed face-on, the projections of a ring of radius $r$ within the disc are circles on the plane of sky. The arrival impact parameter, $b$, of a photon emitted from a radius $r>r_\mathrm{H}$ only depends on $r$ and can be approximated by `just adding one' \citep{gates2020maximum}. This relation, valid for a Kerr black hole, can be rewritten for a Schwarzschild black hole as
\[\frac{b}{M} \approx \frac{1}{2}\left[\frac{r}{M}+1+\sqrt{\left(\frac{r}{M}\right)^2+2\frac{r}{M}-1}\:\right]\approx\frac{r}{M}+1\hspace{0.7em}\text{(face-on)}\:.\tag{3} \label{eq:3}\]
Hence, the event horizon, situated at the radius $r=2M$, appears at an apparent approximated radius $b\approx 2.82M$, in agreement with the numerical outputs of Figure \hyperref[fig:apparent_radii]{2}, while outer radii develop towards higher values of $b$. As for geodesics of order $n=1$, if one stops the backward ray-tracing integration, properly introduced in Section \hyperref[sec:simu]{5}, before getting the third crossing of the plane, the numerical inner and outer edges of the $n=1$ lensing band appear at $b=5.02M$ and $b=6.17M$ respectively \citep{gralla2019black}; these values agree with Figure \hyperref[fig:apparent_radii]{2}.

If the screen and the equatorial plane are not parallel, the images of equatorial rings are deformed by projection effects.\footnote{In a Newtonian context, they would give rise to ellipses.} On the contrary, the critical curve does not vary under change of the plane's inclination, $i$, because of the spherical symmetry of the photon sphere and it always takes the shape of a circle of radius $b=3\sqrt{3}M$ on the plane of sky.\footnote{In Kerr, the photon shell has an equatorial reflection symmetry and the critical curve is always symmetric with respect to the axis perpendicular to the projected direction of the spin \citep{gralla2020lensing}.}\\
\indent Next, we focused on the general properties of projections on the observer's screen of equatorial circles as seen by an inclined observer. The discussions of the following paragraphs are complemented by additional details given in Appendix \hyperref[app:lensing_bands]{A}.
First of all, both the zeroth-order and the first-lensed projections of planar circular rings are equal to their non-inclined counterparts along the line on the screen parallel to the inclination's axis of rotation of the plane (horizontal axis in the upper panels of Figure \hyperref[fig:theoretical_features]{1}). Moreover, projected rings always possess a mirror symmetry on the observer's screen with respect to the axis that is normal to the inclination's one\footnote{This property, reliant on the planar motion typical of all static spherical symmetric space-times, is broken in Kerr \citep{gralla2020lensing}.} (vertical axis in Figure \hyperref[fig:theoretical_features]{1}).\\
\indent Let $\psi$ be the polar angle on the observer's screen as shown in Figure \hyperref[fig:theoretical_features]{1}. If the radius of the inclined ring satisfies $r_\mathrm{H} < r \leq \tilde{r}$, the zeroth-order projection of the side of the ring tilted backwards ($0^\circ < \psi < 180^\circ$, ) or forwards ($180^\circ < \psi < 360^\circ$) appears, respectively, at a higher or lower radial distance on the screen, $b$, with respect to the non-inclined apparent circle. As a result, since in our simulations the forward side of the plane points towards $\psi=270^\circ$, the impact parameter, $b$, of the inclined projected equatorial horizon, as well as that of rings with $r\leq3M$, decreases from $\psi=90^\circ$ to $\psi=270^\circ$ (see grey and magenta lines in the upper panels of Figure \hyperref[fig:apparent_radii]{2}). The same behaviour of $b$ is followed for the forward part of rings with $r>3M$: it then decreases from $\psi=180^\circ$ to $\psi=270^\circ$ in our set-up (see green and yellow lines in the upper panels of Figure \hyperref[fig:apparent_radii]{2}). Instead, for the backward points of rings with $r>3M$, for a given pair of plane's inclination and direction on the observer's screen, there exist a radius for which the behaviour of $b$ is reversed, that is to say that the projection of a given point has a smaller impact parameter than its corresponding non-inclined projection. The existence of this radius, which is smaller for higher inclinations and for directions on the screen closer to the axis of reflection, is a necessary consequence of the flat space-time limit: in a Minkowski space-time, the projected image of a circle is an ellipse whose semi-major axis is along the rotation axis (the horizontal direction here) and equal to the non-inclined projection. Since the semi-minor axis is smaller than the semi-major axis by definition, a reversal must appear at large distance from the black hole to reproduce the Minkowski behaviour.
For the small inclination considered in the set-up of Figure \hyperref[fig:apparent_radii]{2}, this effect appears beyond the considered field of view.
\\
\indent As for the first-lensed projection of equatorial rings, if the radius of the ring satisfies $r_\mathrm{H} < r \leq \tilde{r}$, the projection of the forward or backward side of the ring appears, respectively, at a higher or lower radial distance on the screen, $b$, with respect to the non-inclined apparent circle. Hence, as the geodesic executes a half turn around the black hole before touching the screen and since in our simulations the forward side of the plane points towards $\psi=270^\circ$, the impact parameter, $b$, of such rings decreases from $\psi=90^\circ$ to $\psi=270^\circ$ (see grey and magenta lines in the lower panels of Figure \hyperref[fig:apparent_radii]{2}). If $r=3M$, there is no variation with respect to the non-inclined case (see green line in the lower panels of Figure \hyperref[fig:apparent_radii]{2}). Finally, if $r>3M$, the modulations of $b$ are reversed.
Thus, in our simulations, $b$ increases from $\psi=90^\circ$ to $\psi=270^\circ$ (see yellow line in the lower panels of Figure \hyperref[fig:apparent_radii]{2}). Put differently, the ring projections flee the critical curve in the lower part of the screen and are attracted to it in the upper part.

\subsection{Inner shadow, primary image, and photon rings}
\label{sec:shadow}
The relevant observables appear on the far-away observer's screen if and only if emitting material is present around the black hole. In particular, a single point source gives rise to an infinite sequence of images. These additional lensed observables were originally called `ghost images' \citep{darwin1959gravity} and were present in the first simulated `photograph' of an accretion disc around a Schwarzschild\footnote{The first image of an accretion disc around a Kerr black hole with $a\neq0$ has been computed by Viergutz \citep{viergutz1993image}.} black hole \citep{luminet1979image}. Supermassive black holes such as Sgr A* and M87* are surrounded by emitting matter forming a hot, geometrically thick, and optically thin radiatively inefficient accretion flow \citep[RIAF,][]{yuan2014hot}. Moreover, for M87*, EHT data favours the scenario of a magnetically arrested disc (MAD) whose millimetre wavelength emission arises near the event horizon, predominantly in the vicinity of the equatorial plane \citep{narayan2003magnetically,EHTV,EHTVIII}.\\
\indent Images of black holes surrounded by optically thin emission display a brightness depression in their inner region, known as observable shadow, which does not always track the critical curve, but it is instead highly astrophysical dependent  \citep{chael2021observing}. For instance, for an idealistically thin equatorial disc, this `inner shadow' can be well inside the critical curve: when the disc extends below the photon sphere, the edge of this darkest region corresponds to the projection of the inner radius of the equatorial disc modulated by redshift effects \citep{chael2021observing}. For a disc seen face-on and extending to the event horizon of a Schwarzschild black hole, the inner shadow has a radius of about $2.82M$ on the observer's screen, as discussed in Paragraph \hyperref[par:rings]{2.3}. \\
\indent The primary image is the one reaching the observer on geodesics touching the emitting equatorial source just once. It is also called $n=0$ or direct image. Photon rings are higher-order images of the same emitting material, labelled by the number $n\geq1$ of the equatorial-plane crossings along their nearly bound orbit. When the source is intersected during the crossing, a flux increment is added. Also, the $n$-th photon ring of the equatorial source lies inside the $n$-th lensing band \citep{gralla2020lensing}.\\
\indent When the emitting source is not isotropically distributed, as in the case of equatorial discs, photon rings exponentially tend to the critical curve, have exponentially decreasing angular width and thus are exponentially demagnified and exponentially loose flux density\footnote{The first M87* photon ring provides $\sim10\%$ of the total luminosity in numerical GRMHD simulations \citep{johnson2020universal}.} for increasing $n$ \citep{johnson2020universal,vincent2022images}. Also, in intensity cross sections of the images, they do not blend with the direct image, unlike an infalling sphere, but they give birth to a stack of discrete higher-order images superimposed on top of the primary one \citep{gralla2019black,vincent2022images}, provided that absorption effects are weak at a given observing frequency \citep{beckwith2005extreme}. These properties are presented more meticulously in the introduction of \citet{paugnat2022photon}.\\
\indent To summarise, photon rings are signatures related to the strong-field lensing effects of the photon sphere and, unlike the critical curve or the lensing bands, they can be observed and possess only limited astrophysical dependence \citep{vincent2022images}. The present EHT does not resolve the photon rings \citep{gralla2021M87}, but their interferometric detection may soon become possible by observing at higher frequencies \citep[ngEHT,][]{ngEHT} or through longer Earth-space baselines \citep[BHEX,][]{BHEX}.

\section{Compact object model}
\label{sec:metric}

\subsection{M87* distance and orientation}
Our compact object model mimics the observationally assessed properties of M87*, namely its mass, $M$, its distance, $D$, and its inclination, $i$. The values of $M$ and $D$ are those inferred by the EHT collaboration \citep{EHTVI}, the mass being consistent with that obtained via a stellar-dynamics study \citep{gebhardt2011black}. For its inclination, defined as the angle between the line of sight and the angular momentum of the accretion flow, we used the value of the viewing angle of the jet in M87~\citep{mertens2016kinematics}, assuming that its axis is aligned with the angular momentum vector of the disc and that the latter points away from the observer. 

\subsection{Rezzolla-Zhidenko metric}
In this paper, we consider the Rezzolla-Zhidenko metric \citep[RZ metric,][]{rezzolla2014new}, which describes in a parametric framework the space-time outside a general spherically symmetric,\footnote{The Konoplya-Rezzolla-Zhidenko metric \citep[KRZ metric,][]{konoplya2016general} generalises the RZ parametrisation to circular space-times, then to rotating black holes, but not all stationary axisymmetric black holes are included in this class.} stationary, and therefore static black hole.\footnote{\citet{kocherlakota2020accurate} extend the RZ scheme to arbitrary asymptotically flat, spherical symmetric, and static space-times, including non--black hole cases such as boson stars or naked singularities.} In the strong-field regime, contrary to previous parametrisations in which an infinite number of equally important parameters is needed \citep{johannsen2011metric}, the RZ metric is described by a finite number of hierarchical factors. The line element of this metric in a spherical polar coordinate system $(t,r,\theta,\varphi)$ is
\[ ds^2 = - N^2(r)dt^2+\frac{B^2(r)}{N^2(r)}dr^2+r^2(d\theta^2+\sin^2\theta d\varphi^2)\:, \tag{4} \label{eq:4}\] 
where the functions $N$ and $B$ only depend on the radial coordinate, $r$. Conveniently for a black hole, $r$ can be compactified in the dimensionless variable $x\coloneqq1-r_\mathrm{H}/r$, where $r_\mathrm{H}$ is the radial position of the event horizon so that $x=0$ corresponds to the location of the horizon and $x=1$ to spatial infinity. In terms of $x$, using $G=c=1$ natural units and introducing $\epsilon\coloneqq2M/r_{\mathrm{H}}-1$, the functions $N$ and $B$ can be written as follows:
\begin{equation*}
    \begin{cases}
        N^2(x) = x [1-\epsilon(1-x)+(a_0-\epsilon)(1-x)^2+\tilde{A}(x)(1-x)^3]\\
        B(x) = 1+b_0(1-x)+\tilde{B}(x)(1-x)^2
\end{cases}\hspace{-0.5cm}\tag{5} \label{eq:5}\:,
\end{equation*}
where $\tilde{A}$ and $\tilde{B}$ are approximated in the form of the following continued fractions:
\begin{equation*}
    \tilde{A}(x) = \frac{a_1}{1+\dfrac{a_2\,x}{1+\dfrac{a_3\,x}{1+\ldots}}}\quad\color{black},\quad 
    \tilde{B}(x) = \frac{b_1}{1+\dfrac{b_2\,x}{1+\dfrac{b_3\,x}{1+\ldots}}}\:\:\,.\tag{6} \label{eq:6}
\end{equation*}
The dimensionless constants $a_0,a_1,a_2,a_3\ldots$ and $b_0,b_1,b_2,b_3\ldots$ have to be constrained by observations. If all $\epsilon$, $a_i$, and $b_i$ are zero, with $i\in \mathbb{N}$, the Schwarzschild metric is recovered in Schwarzschild-Droste coordinates.\\
\indent It is useful to notice that at the horizon the functions $\tilde{A}$, $\tilde{B}$ of equations \eqref{eq:6}, reduce to $\tilde{A}(0)=a_1$, $\tilde{B}(0)=b_1$.
This means that near the horizon, only the lower-order terms in the expansion are relevant. Higher-order terms increase the accuracy of the approximation and, for numerous known metrics deviating slightly from the Schwarzschild one, the first order is sufficient to accurately investigate astrophysical observable deviations \citep{konoplya2020general}. Besides, the coefficients $a_0$ and $b_0$, which have been introduced in the equations \eqref{eq:5}, can be expressed as a combination of the parametrised post-Newtonian (PPN) parameters and, assuming black hole uniqueness, they are constrained by observations to be at most of the order of $10^{-4}$ \citep{will2014confrontation}.\\
\indent For the sake of simplicity, as we were interested in the near-horizon emission around a black hole and following the arguments stated before, in the simulations all the parameters $a_i$, $b_i$ with $i\geq 2$ were put to zero. 
The parameters $a_0$ and $b_0$ were also fixed to be zero, their impact on the image formation being negligible with respect to that of the astrophysical set-up when PPN constraints are met, as it is illustrated in Appendix \hyperref[app:parameters]{C}; in the same Appendix we also present higher deviations in $a_0$ and $b_0$, allowed in theories of gravity without a Birkhoff-like uniqueness theorem.
Finally, we chose $b_1=0$, again for simplicity purposes and because its variation has a very limited effect on the image, as shown in Section \hyperref[sec:analysis]{6} and Appendix \hyperref[app:parameters]{C}. We could then study the effect on the images of the only unconstrained parameters, $a_1$ and $\epsilon$, which we made vary 
independently by setting the other parameter to zero. We chose them in their theoretically allowed\footnote{The outermost horizon is located at $r=r_\mathrm{H}$ and no larger roots of $g_{rr}^{-1}=N^2(r)$ should exist \citep{kocherlakota2022distinguishing}.} ranges and took `small', in the sense of the following right inequality,\footnote{This does not imply that the deviations are physically small.} parametric deviations \citep{kocherlakota2020accurate}:
\[-1<\epsilon\leq0.5, \quad a_1\geq-1\quad\text{and}\quad \max{(\,|\,a_1\,|,|\,\epsilon\,|\,)} \leq 1\:. \tag{7} \label{eq:7}\] 
\begin{table}[!ht]
\caption{Parameters of the black hole.}            
\label{tab:BH}      
\centering                    
\begin{tabular}{c c c}        
\hline\hline                 
Symbol & Value & Property\\    
\hline                        
   $M$ & $6.5\times10^9$ $\mathrm{M}_{\odot}$& Mass \\
   $D$ & 16.9 Mpc & Distance \\      
   $i$ &  163° & Inclination\\
   $a_0$, $b_0$, $b_1$, $a_i$, $b_i$ ($i\geq2$) & 0 & RZ parameters \\
   $a_1|_{\,\epsilon=0}$  & $\{0,\:0.1\}$ & RZ parameter \\
   $\epsilon\,|_{\,a_1=0}$  & $\{0,\:0.1\}$ & RZ parameter \\
\hline                        
\end{tabular}
\end{table}

\section{Accretion disc model}
\label{sec:disc}

In this paper, the compact object is surrounded by an infinitely geometrically thin accretion disc orbiting in the equatorial plane taken to be at $\theta=\pi/2$ without any loss of generality. Our disc extends from an inner radius, $r_\mathrm{inner}$, given by the radial position of the event horizon, to a sufficiently large outer radius, $r_\mathrm{outer}$, whose contribution to the image formation is negligible for our choice of quickly decreasing emission profiles. 

\subsection{Motion of the emitting material}
\begin{figure}[!t]
\label{fig:velocity}
    \centering
\includegraphics[width=\hsize]{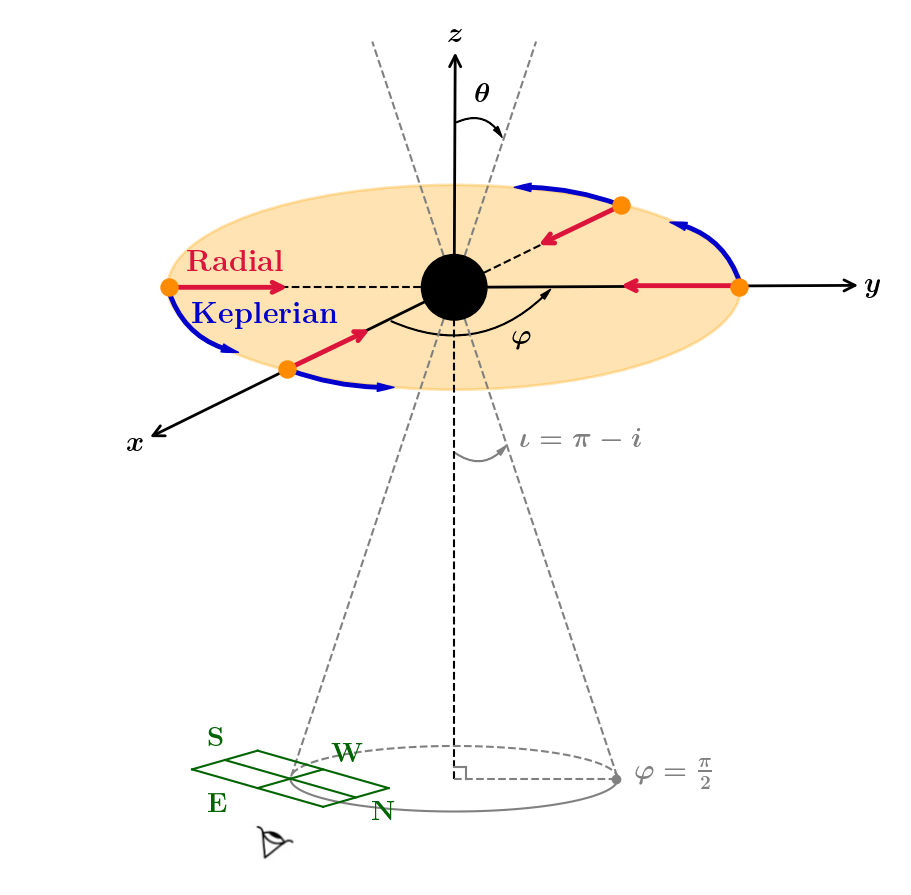}
    \caption{Illustration of the velocity fields and of the ray-tracing set-up. The orange massive particles of the yellow accretion disc can have a blue Keplerian velocity, be radially infalling along the red vectors or possessing a mixed velocity combining these two extreme cases. Schwarzschild-Droste coordinates are depicted in black, while the grey quantities locate the green screen of the observer.}
    \label{fig:enter-label}
\end{figure}

We supposed that the emitters of mass $m_\mathrm{e}>0$ can have three possible types of dynamics: the particles in the disc rotate around the black hole with a Keplerian velocity, they fall radially towards the central singularity, or they possess a more realistic mixed velocity. A sketch of the two extreme velocity fields is reproduced in Figure \hyperref[fig:velocity]{3}. We defined the specific energy, $\varepsilon$, and specific angular momentum, $\ell$, as
\[\varepsilon\coloneqq\frac{E}{m_\mathrm{e}}\quad\text{and}\quad\ell\coloneqq\frac{L}{m_\mathrm{e}}\:.\tag{8} \label{eq:8}\]
\subsubsection{Keplerian velocity}
\label{sec:keplerian}

If the disc is Keplerian, the accreted matter follows a circular geodesic motion above the innermost stable circular orbit (ISCO), below which we assumed that it spirals down along time-like geodesics with constants of motion, $\varepsilon$ and $\ell$, equal to those of the ISCO \citep{cunningham1975effects}. A general time-like 4-velocity, $u^{\,\mu}=p^{\,\mu}/m_\mathrm{e}=(u^t,u^r,0,u^\varphi)$, can be expressed with the conserved quantities and via the normalisation $u^{\,\mu} u_\mu=-1$. Noting by primes the derivatives with respect to $r$, the expressions of the specific energy and z-component of the angular momentum for the RZ metric are then \citep{cardenas2019experimental}
\[\varepsilon_\mathrm{circ}=\sqrt{\frac{N^3(r)}{N(r)-rN'(r)}}\quad\text{and}\quad\ell_\mathrm{circ}=\sqrt{\frac{r^3N'(r)}{N(r)-rN'(r)}}\:,\tag{9} \label{eq:9}\]
and the disc circular orbiters satisfy two additional conditions on the effective potential of the radial geodesic motion:
\[ V_{\mathrm{eff}} = 0 = V'_{\mathrm{eff}}\quad \text{with} \quad V_{\mathrm{eff}} = \frac{\varepsilon_\mathrm{circ}^2}{N^2(r)}-\frac{\ell_\mathrm{circ}^2}{r^2}-1\:.
\tag{10} \label{eq:10}\]
The last stable orbit being marginally stable, the second derivative of the effective potential, $V''_{\mathrm{eff}}$, must vanish at the radius of the ISCO, $r_\mathrm{ms}$, which is then the real root of the following  equation with no analytical solution \citep{rezzolla2014new}:
\[3N(r_\mathrm{ms})N'(r_\mathrm{ms})-3rN'^2(r_\mathrm{ms})+rN(r_\mathrm{ms})N''(r_\mathrm{ms})=0\:.
\tag{11} \label{eq:11}\]

\subsubsection{Radial infall}
\label{sec:radial}

When the accreted matter is in radial free-fall towards the central compact object, the particle evolves in a constant-$\theta$ plane with vanishing specific angular momentum, $\ell$. The conservation of the specific energy, $\varepsilon$, and the normalisation of the 4-velocity lead to
\citep{kocherlakota2022distinguishing}:
\[u^\mu_\mathrm{rad}=\left(\frac{\varepsilon}{N^2},-\frac{\sqrt{\varepsilon^2-N^2}}{|\,B\,|},0,0\right)\:.
\tag{12} \label{eq:12}\]
We also neglected the radial velocity at infinity so that
\[ \lim_{r\to\infty} u^r_\mathrm{rad} = 0 \quad \text{and}\quad \varepsilon>0 \quad \implies \quad  \varepsilon=1\:.\tag{13} \label{eq:13}\]

\subsubsection{Mixed velocity}
\label{sec:mixed}
Realistic RIAF discs are neither totally radially infalling nor Keplerian, and according to more sophisticated GRMHD simulations \citep{narayan2012grmhd}, they likely consist of a mixed velocity comprising a radial component and a sub-Keplerian part \citep{yuan2022accretion,begelman2022really}. Moreover, the sub-Keplerian motion is expected to be predominant for the disc surrounding M87* \citep{chatterjee2023accretion}.

To define the sub-Keplerian velocity, $u_{\mathrm{sub}}$, we followed the same reasoning of Paragraph \hyperref[sec:keplerian]{4.1.1}, but we multiplied the quantity $\ell/\varepsilon$ by a `sub-Keplerianity' factor, $\xi\in(0,1]$. Thus, above the ISCO 
\[\varepsilon_\mathrm{sub}=\sqrt{\frac{N^3(r)}{N(r)-\xi^2rN'(r)}}\:,\quad\ell_\mathrm{sub}=\xi\sqrt{\frac{r^3N'(r)}{N(r)-\xi^2rN'(r)}}\:,\tag{14} \label{eq:14}\]
whereas below the ISCO $\varepsilon_\mathrm{sub}=\varepsilon_\mathrm{sub}(r_\mathrm{ms})$ and $\ell_\mathrm{sub}=\ell_\mathrm{sub}(r_\mathrm{ms})$.
Then, introducing the angular velocity, $\Omega=u^\varphi/u^t$,\footnote{In spherical symmetric space-times, $\Omega_\mathrm{rad}$ is always zero.} and the parameters $0\leq\omega_r\leq1$ and $0\leq\omega_\varphi\leq1$, we linearly combined the sub-Keplerian and radial 4-velocities \citep{pu2016effects,cardenas2023adaptive}:
\begin{align*}
    u_\mathrm{mix}^r &= u_\mathrm{sub}^r+\left(1-\omega_r\right)\left(u_\mathrm{rad}^r-u_\mathrm{sub}^r\right)\:,
    \tag{15.1} \label{eq:15.1}\\[4pt]
    \Omega_\mathrm{mix} &=\Omega_\mathrm{sub}+\left(1-\omega_\varphi\right)\left(\Omega_\mathrm{rad}-\Omega_\mathrm{sub}\right)\:.
    \tag{15.2} \label{eq:15.2}
\end{align*}
Finally, $u^t_\mathrm{mix}$ was found by unit-normalisation and $u^\varphi_\mathrm{mix}=\Omega_\mathrm{mix}u^t_\mathrm{mix}$. 

We note that this realistic  mixed velocity does not correspond to a geodesic motion contrary to the purely Keplerian and radial ones, which are retrieved, respectively, by imposing $\omega_r=\omega_\varphi=1$ with $\xi=1$ and $\omega_r=\omega_\varphi=0$. In other words, the mixed velocity prescription can account for non-gravitational forces, which are expected to affect the emitters' trajectory. 

\subsection{Synchrotron radiation}

We assumed the disc to be optically and geometrically thin, which means considering the optical depth to be idealistically null. The first hypothesis is only an approximation for the present EHT images of M87* and particularly for the photon rings observations, for which the opacity effects may play a role \citep{vincent2022images}. The assumption holds much better at 345 GHz, as absorption effects decrease with increasing frequency, and so it is reasonable for the observations at higher frequency of the ngEHT \citep{doeleman2023reference} or BHEX \citep{BHEX}. 
In the absence of absorption, the quantity $I_\nu/\nu^3$, with $I_\nu$ the intensity per unit frequency, is conserved along the geodesic \citep{lindquist} so that the source specific intensity, $I_\nu^\mathrm{em}$, is related to the observed intensity, $I_\nu^\mathrm{obs}$:
\[ \frac{I_\nu^\mathrm{obs}}{I_\nu^\mathrm{em}} \propto \left(\frac{\nu^\mathrm{obs}}{\nu^\mathrm{em}}\right)^3, \tag{16} \label{eq:16}\]
where $\nu^{\mathrm{obs}/\mathrm{em}}$ is the observed or emitted frequency. Also, the quantity between parentheses represents the redshift factor:
\[ g\coloneqq\frac{\nu^\mathrm{obs}}{\nu^\mathrm{em}}=\frac{\boldsymbol{p}^\mathrm{obs}\cdot \boldsymbol{u}^\mathrm{obs}}{\boldsymbol{p}^\mathrm{em}\cdot \boldsymbol{u}^\mathrm{em}}\:,\tag{17} \label{eq:17}\]
defined as the scalar product between the photon momentum at the observer or emitter, $\boldsymbol{p}^{\mathrm{obs}/\mathrm{em}}$, and the 4-velocity of the observer or emitter, $\boldsymbol{u}^{\mathrm{obs}/\mathrm{em}}$.\\
\indent Even though plasma-physics phenomena, such as magnetic reconnection, can accelerate a population of electrons in power-law distributions, we assumed, to simplify, that the entire disc emits relativistic thermal synchrotron radiation; which is reasonable to characterise M87* emission at millimetre wavelengths.\footnote{The assumption is not appropriate for the infra-red flaring of Sgr A*.}\\
\indent Thus, according to equation (B3) in \citet{vincent2022images}, and making the same approximations detailed therein in Appendix A, the relativistic electrons follow the Maxwell-Jüttner distribution at temperature $T_\mathrm{e}$, whose emissivity in cgs units is
\begin{align*}
  j_\nu &\approx \frac{\sqrt{2}\pi e^2}{6c}\frac{n_\mathrm{e}\nu_{\mathrm{em}}}{\Theta_{\mathrm{e}}^2}
  \exp\left[-\left(\frac{9\pi m_\mathrm{e} c}{e\sin\theta_\mathrm{B}}\frac{\nu_\mathrm{em}}{B\Theta_{\mathrm{e}}^2}\right)^{1/3}\right],\tag{18} \label{eq:18}\\[4pt]
  \theta_\mathrm{B}&=\arccos(\boldsymbol{\bar{K}}\cdot\boldsymbol{\bar{B}})
  \:,\tag{18.1} \label{eq:18.1}	
\end{align*}
where $n_\mathrm{e}$ is the number density of the electrons of mass $m_\mathrm{e}$, $e$ the charge of the electrons, $B$ the amplitude of the magnetic field, $\theta_\mathrm{B}$ the pitch angle between the unit vector along the direction of emission, $\boldsymbol{\bar{K}}$, and the unit magnetic field vector, $\boldsymbol{\bar{B}}$, both measured in the comoving frame of the emitting plasma,\footnote{A rigorous definition of $\boldsymbol{\bar{K}}$ and $\boldsymbol{\bar{B}}$ is given in Appendix \hyperref[app:theta_mag]{C.3}.} and $\Theta_\mathrm{e}\coloneqq k_\mathrm{B}T_\mathrm{e}/m_\mathrm{e}c^2$, $k_B$ being the Boltzmann constant.\\
Because the disc of our simulation is geometrically and optically thin, the radiative transfer equation reduces to 
\[I_\nu^\mathrm{em} \propto j_\nu\:.\tag{19} \label{eq:19}\]
\indent In order to study the effects of the emission process on the simulated images, we assumed that the physical quantities, $n_\mathrm{e}$, $\Theta_\mathrm{e}$, and $B$, follow power-law radial distributions with indices $\alpha$, $\beta$, and $\gamma$ varying in ranges including the values obtained by 3D GRMHD simulations \citep{Cho,chatterjee2023accretion} and in fair agreement with the ones set by \citet{desire2025multifrequency}:
\begin{gather*}
 \begin{aligned}
n_e&\propto r^{-\alpha},\quad \alpha \in [1;1.5] \\ \Theta_e&\propto r^{-\beta}, \quad
\beta \,\in [1;2]\\
B&\propto r^{-\gamma}, \quad
\gamma =1\:.   
 \end{aligned}\tag{20} \label{eq:20}
\end{gather*} 
Substituting these power-laws in the emissivity, introducing the indices $i_1=\alpha-2\beta$ and $i_2=\gamma+2\beta$, gives
\begin{align*}
j_\nu &\approx \eta \frac{\nu_\mathrm{em}\text{\scriptsize[GHz]}}{230}
\left(\frac{r}{r_\mathrm{inner}}\right)^{-i_1}
\hspace{-0.1cm}\exp\left[-\zeta
\sqrt[3]{\frac{\nu_\mathrm{em}\text{\scriptsize[GHz]}}{230\sin\theta_\mathrm{B}}}
\left(\frac{r}{r_\mathrm{inner}}\right)^{i_2/3}
\right], \tag{21} \label{eq:21}\\[4pt]
  \eta &\approx 1.31\times10^{-18}\,
\frac{n_{\mathrm{e};\mathrm{inner}}}{\Theta_{\mathrm{e;inner}}^2}\approx 10^{-3}\:\,\text{\footnotesize Jy}\:\text{\footnotesize cm}^{-1}\:, \tag{21.1} \label{eq:21.1}\\[4pt]
\zeta&=\,
\left(\frac{3.7\times10^{5}}{ B_\mathrm{inner}\Theta_{\mathrm{e;inner}}^2}\right)^{1/3}
\in[2.5;4]\:.\tag{21.2} \label{eq:21.2}
\end{align*}
\indent The value of $\zeta$, justified in detail in Appendix B of \citet{vincent2022images}, was chosen in order to match the predicted characteristics of the accretion flow around M87*. The normalisation, $\eta>0$, ensures that the 230 GHz radiative flux density is of the order of 0.5 Jy for a mean profile ($\zeta$=3, $\alpha$=1, $\beta$=2) as suggested by the EHT data \citep{EHTIV,wielgus2020monitoring,akiyama2024persistent}. Hence, our simulations only result in a correct flux density for this particular set of parameters. Nevertheless, since the normalisation does not influence the conclusions of this work, we kept the value of $\eta$ constant for simplicity.

\begin{table}[ht!]
\caption{Parameters of the accretion disc.}             
\label{tab:astro}      
\centering                    
\begin{tabular}{c c c}        
\hline\hline                 
Symbol & Value & Property \\   
\hline                        
   $\xi$ & 0.7 & Sub-Keplerian parameter \\
   $\omega_\mathrm{r}$, $\omega_\mathrm{\varphi}$ & 0.8 & Mixed-velocity parameters \\ 
   $\zeta$ & \{2.5,\:3,\:4\} & Emission parameter \\      
   $\alpha$ & \{1,\:1.5\} & Density power-law index\\
   $\beta$ & \{1,\:2\} & Temperature power-law index \\
   $\gamma$ & 1 & Magnetic field power-law index \\
   $\eta$ & $10^{-3}$ & Normalisation factor \\ 
   $r_{\mathrm{inner}}$ & $r_H$ & Inner radius\\
   $r_{\mathrm{outer}}$ & 50 M & Outer radius \\
\hline                        
\end{tabular}
\tablefoot{Fiducial parameters of a mean profile: $\zeta$=3, $\alpha$=1, $\beta$=2.}
\end{table}

\section{Adaptive backward ray-tracing}
\label{sec:simu}

\begin{figure}[ht!]
   \centering
   \includegraphics[width=\hsize]{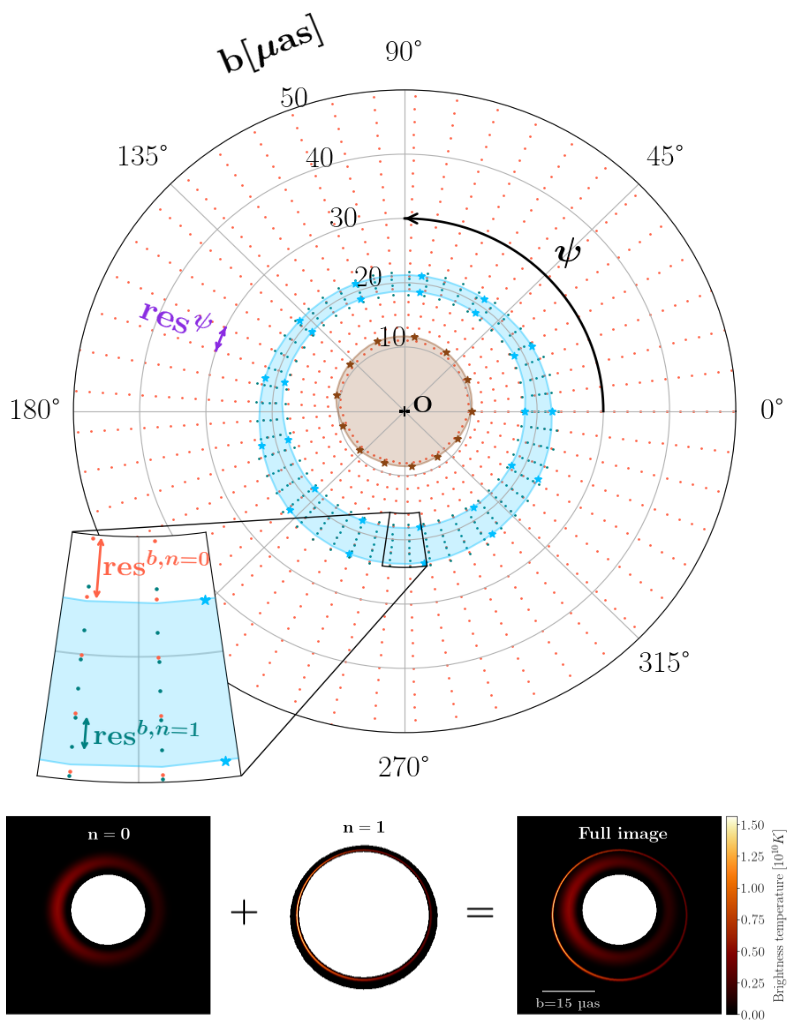}
      \caption{Adaptative ray-tracing polar grid with point densities adjusted for visual purposes. The real resolution values used in the simulated images are: angular resolution $\mathrm{res}^{\psi}=3.6^{\circ}$ (violet), $n=0$ radial resolution $\mathrm{res}^{b,\:n=0}=0.05$ µas (orange) outside the inner shadow (brown), $n=1$ radial resolution $\mathrm{res}^{b,\:n=1}=0.005$ µas (teal) in its lensing band (light blue). The star symbols represent the computed points of the apparent horizon and of the edges of the $n=1$ lensing band before interpolation. The inner 30 µas of a mean-profile sample images are also shown.}
\label{fig:adaptative}
\end{figure}

To simulate our images we used the ray-tracing code GYOTO \citep[General RelativitY Orbit Tracer of the Observatoire de Paris,][]{vincent2011gyoto}, collecting the specific intensity, $I_\nu$, along null geodesics that are shot from pixels making up the screen of a distant observer towards the compact object. The resulting image is then a map of $I_\nu$ on the observer sky.\\
\indent Keeping track of the equatorial crossings, it is possible to decompose the simulated image into different $n$ orders, properly defined for light rays in Section \hyperref[sec:img]{2}: the $n=0$ image is the map of the specific intensity obtained along the part of geodesic integrated before the second equatorial plane encounter, while the $n=1$ image is the map of $I_\nu$ accumulated between the second and the third crossing. We restricted our analyses to the zeroth and first orders so that the geodesic integration leading to the $n\geq 2$ images was not performed.\\
\indent We defined a polar coordinate system on the observer's screen, that is, on the 2D simulated image, so that every point on it is described by its distance, $b$, from the centre, $O$, and by the positive angle, $\psi$, from the horizontal axis passing from $O$ (see Figure \hyperref[fig:adaptative]{4}).\\
\indent For each metric of Section \hyperref[sec:metric]{3}, we produced adaptively ray-traced images based on the identification of the projection of the equatorial horizon and lensing bands introduced in Section \hyperref[sec:img]{2}, adapting the resolution for different image orders which are characterised by different scales. As explained in \citet{gelles2021role}, this is an efficient way to explore the properties of computationally expensive photon rings. To do that, we first simulated the image of a disc with constant emission and without spectral redshift effects ($\nu_\mathrm{em}=\nu_\mathrm{obs}$, $\zeta=0$ and $i_1=0$) on pixels separated by 0.1 µas along 14 equally spaced polar angular directions on sky. The disc was taken to have a very large effective outer radius in this case. Then, since the order of a null geodesic is related to the number of equatorial crossings, the structure of the image could be distinguished thanks to the number of impacts of the photon with the disc, itself linked to the value of $I_\nu$. In other words, for each angular direction, the boundary of the projected horizon was given by the further pixels from the centre having a null intensity, as no emission originates from the interior of the black hole by definition, and the edges of the $n=1$ lensing band were defined by the closest and furthest pixels cumulating the maximum intensity, since their geodesic collected twice the constant emission of the disc. Afterwards, the 14 computed values on the edges were connected with an interpolated curve. Next, for the actual emission profiles of Section \hyperref[sec:disc]{4}, the $n=0$ image, computed for the pixels in the mask extending outside the apparent horizon until $r_\mathrm{outer}$ and the $n=1$ image, computed only inside the mask corresponding to the $n=1$ lensing band, with margins of 0.1 µas around the desired zone in order to be sure not to lose some needed pixels because of a bad initial resolution, were traced with adaptively chosen resolutions, $\mathrm{res}^{b,\:n=0,1}$. \\
\indent Our images were simulated at the two\footnote{At 86 GHz absorption effects cannot be neglected.} highest observing frequencies of the ngEHT and BHEX, 230 and 345 GHz \citep{doeleman2023reference,BHEX}, and the screen orientation was chosen in such a way that the direction of the projected angular-momentum vector of the disc is aligned with the vertical axis.
The field of view and the integration variables were chosen in order to have a reasonable compromise between a good image resolution and a modest computing time. The parameters of our simulations are given in Tables \hyperref[tab:BH]{1}, \hyperref[tab:astro]{2} and \hyperref[tab:simu]{3}.

\begin{table}[ht!]
\caption{Parameters of the simulation.}             
\label{tab:simu}      
\centering                     
\begin{tabular}{c c c}       
\hline\hline           
Symbol & Value & Property\\    
\hline                        
   $\nu^\mathrm{obs}$ & \{230,\:345\} GHz & Observing frequency\\ 
   $\mathrm{fov}$ & 100 µas & Field of view\\
   $\mathrm{res}^{\psi}$ &3.6° & Angular resolution\\
   $\mathrm{res}^{b,\:n=0}$ & 0.05 µas & $n=0$ radial resolution\\
   $\mathrm{res}^{b,\:n=1}$ & 0.005 µas & $n=1$ radial resolution\\
\hline                         
\end{tabular}
\end{table}

\begin{figure*}[ht!]
   \resizebox{\hsize}{!}
    {\includegraphics{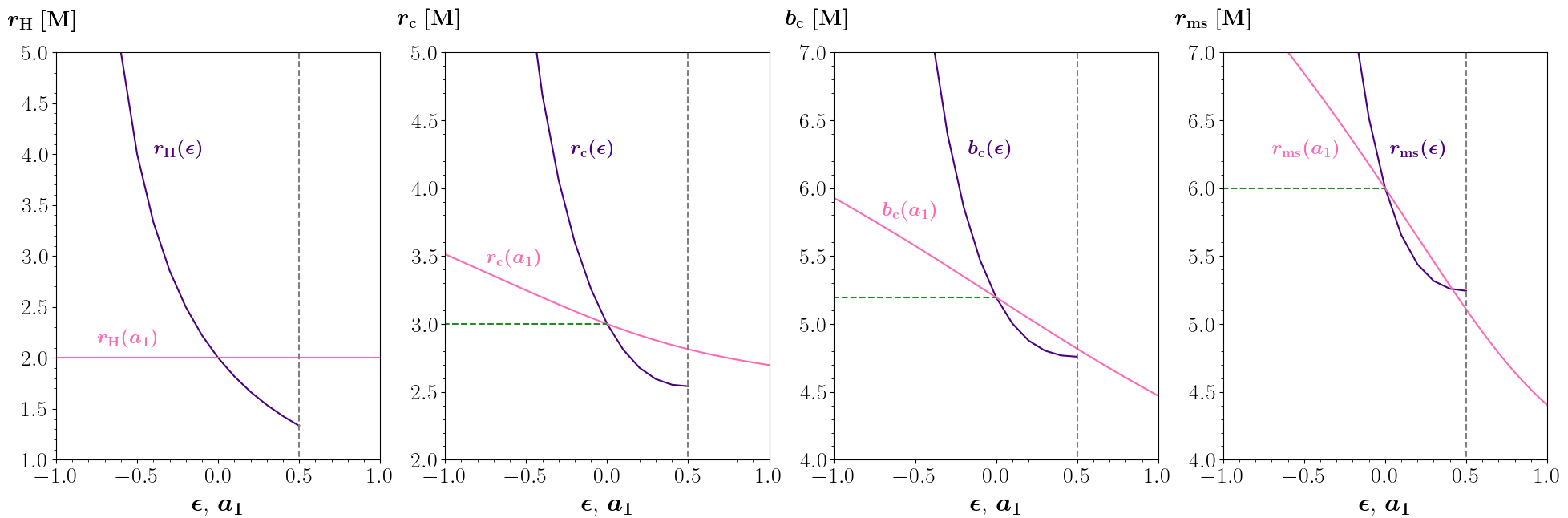}}
    \caption{Dependence on the metric parameters $\epsilon$, $a_1$ of the event horizon, the photon sphere, the critical curve, and the ISCO. The curves depending on $\epsilon$ or $a_1$ are traced at $a_1=0$ or $\epsilon=0$ respectively. Green lines indicate corresponding radii in a Schwarzschild space-time.}
\label{fig:Parameters}
\end{figure*}

\section{Image analysis}
\label{sec:analysis}

Here, we consider an actual intensity map on the observer's screen. We performed 100 1D intensity cuts equally spaced along the angular direction, $\psi$. On each of them, with the same radial resolutions used to ray-trace the simulated primary image and $n=1$ photon ring, we computed the positions of the peaks and the widths of the profiles which are defined hereafter.

\subsection{Position of the peak and width of an intensity profile}

The peak along a given angular direction is the maximum of the 1D intensity profile. So, the location of the peak for a given $\psi$ is the impact parameter, $b$, corresponding to the maximum intensity. Hereafter, we list the footprints on the intensity peak position of different astrophysical configurations or space-time geometries, gradually incorporating complexities such as emission anisotropy and redshift effects.\\
\indent If $\theta_\mathrm{B}$ is fixed to a constant value, the disc emission towards the observer is axisymmetric.\footnote{The local physics of the disc always remains axisymmetric.} Under this simplification, the emissivity, $j_\nu$, of equation \eqref{eq:21} and consequently the emitted intensity, $I_\nu^\mathrm{em}$, related to $j_\nu$ via equation \eqref{eq:19}, reach their maximum at a radius, $r_\mathrm{max}$, computed as
\[ r_\mathrm{max}=-
\left(\frac{3i_1}{i_2\zeta}\sqrt[3]{\frac{230\sin\theta_\mathrm{B}}{\nu_\mathrm{em}\text{\scriptsize[GHz]}}}\right)^{3/i_2}r_\mathrm{H}\:. \tag{22} \label{eq:22}\]
Given the range of parameters of Table \hyperref[tab:astro]{2}, $r_\mathrm{max}\leq r_\mathrm{H}$ and $j_\nu$ is monotonically decreasing after $r_\mathrm{max}$. Therefore, all our emission profiles with $\theta_\mathrm{B}=\mathrm{const}$ radially decrease after $r_\mathrm{H}$. To put it another way, if we neglect frequency shift, that is, we put $g=1$ in equation \eqref{eq:17}, the observed intensity profiles, $I_\nu^\mathrm{obs}$, are equal to the emitted ones, via equation \eqref{eq:16}, and peak, in this case, at the projected radial position of the event horizon for the $n=0$ emission and at the inner boundary of the $n=1$ lensing band, that is the $n=1$ image of the equatorial horizon as explained in Section \hyperref[sec:img]{2}, for the $n=1$ photon ring. Thus, without redshift effects, our constant-$\theta_\mathrm{B}$ observed peak positions can be predicted with the following ingredients:
\begin{itemize}
    \setlength\itemsep{0.1cm}
    \item The metric, which defines the event horizon, the photon shell, and the geodesic motion, and consequently the related image features discussed in Section \hyperref[sec:img]{2}; 
    \item the inclination of the source, determining the deformations due to projection effects (see Section \hyperref[sec:img]{2} and Appendix \hyperref[app:lensing_bands]{A}).
\end{itemize} 
For instance, in the case of a Schwarzschild black hole, the non-redshifted observed peak positions with $\theta_\mathrm{B}$ constant appear to be superposed on the solid, for $n=0$, or dashed, for $n=1$, grey lines in the plots of Figure \hyperref[fig:apparent_radii]{2} and they do not vary under change of the emission profile nor of the velocity of the disc.\\
\indent Still in the non-redshifted framework, if the real value of the pitch angle, $\theta_\mathrm{B}$, between the magnetic field vector and the emission direction in the rest frame of the emitter, is included in the computation of the emissivity, $j_\nu$, of equation \eqref{eq:21}, the prediction of $r_\mathrm{max}$ is less straightforward and the axisymmetry of the emission profile is broken when the disc is not observed face-on; according to equation \eqref{eq:18.1}, the quantities needed to numerically compute $\theta_\mathrm{B}$ are then:
\begin{itemize}
    \setlength\itemsep{0.1cm}
    \item the emitted tangent 4-vector to the photon's geodesic which depends both on the metric and the observer's inclination (this results from lensing effects due to the space-time curvature);
    \item the direction of the magnetic field, taken to be everywhere vertical in our simulations. This assumption is defensible for a preliminary study of the compact emission from a MAD system, where the presence of a strong poloidal magnetic field is favoured \citep{narayan2003magnetically,EHTVIII}. Some results obtained considering the other component of a poloidal structure, namely, a radial magnetic field, are also shown in Appendix \hyperref[app:parameters]{C}. 
\end{itemize} 
\indent \indent If spectral effects are taken into account, the observed intensity, $I_\nu^\mathrm{obs}$, is computed according to equations \eqref{eq:16} and \eqref{eq:17}, so that it is both geometry-reliant and astrophysics-dependant:
\begin{itemize}
    \setlength\itemsep{0.1cm}
    \item the resulting $I_\nu^\mathrm{obs}$ is of course based on the shape of $I_\nu^\mathrm{em}$;
    \item different 4-velocities of the emitter lead to different $g$ factors (this includes both the dynamical, or special relativistic, and the general relativistic redshift effects);
	    \item the emitted or observed 4-momentum of the photon depend on the metric and on the observer's inclination (this is again the lensing effect due to the space-time curvature).
\end{itemize} 
As a result, the peak positions of our observed intensity profiles are shifted with respect to their corresponding non-redshifted impact parameters and this shift depends on the metric as well as on the properties of the accretion disc. Hence, already in the case of a Schwarzschild black hole with $\theta_\mathrm{B}$ fixed, the observed radii of the brightest parts of the images are no more superposed to the direct or lensed projections of the horizon, grey lines of Figure \hyperref[fig:apparent_radii]{2}, and their values do vary under change of the emission profile and of the velocity of the disc.\\
\indent We defined the width of an intensity profile as its full width at half maximum, taking the difference between the impact parameters, above and below the peak's one, whose associated intensity is closer to that of the maximum intensity divided by two.\\
\indent The comments hereinabove regarding the dependencies of the redshift factor shaping the observed intensity, $I_\nu^\mathrm{obs}$, are clearly still valid and relevant when considering its width, but contrary to the peak position which, at least for our class of synchrotron profiles and taking a constant $\theta_\mathrm{B}$, is invariant under change of the emission profile when spectral effects are neglected, the width is intrinsically dependent on the radiative properties of the disc.

\subsection{Impact of the metric on the image features}

We have shown in Sections \hyperref[sec:img]{2} and \hyperref[sec:analysis]{6} that the horizon, the critical curve, and the lensing bands are key elements to interpret black hole images, even though they are purely geometric, non-observable features. For instance, if we do not incorporate lensing effects in the emissivity nor frequency shift, $g$, the peak positions of all our emission profiles track the apparent horizon for the $n=0$ image and the inner boundary of the first lensing band for the $n=1$ photon ring. In order to evaluate the geometrical effect on the image formation, it is then fundamental to understand how these characteristics vary when varying the space-time metric.\\
\indent We recall that the location of the event horizon under the RZ parametrisation \citep{rezzolla2014new} is only governed by the parameter $\epsilon$ and it is given by
\[ r_\mathrm{H} = \frac{2M}{\epsilon+1}\tag{23} \label{eq:23}\:.\]
Thus, $r_\mathrm{H}(\epsilon)$ is a positive decreasing function for $-1<\epsilon$ and the Schwarzschild event horizon, $r_\mathrm{H}=2M$, is retrieved for $\epsilon=0$, as it is shown in the first panel of Figure \hyperref[fig:Parameters]{5}.\\
\indent The radius of the critical curve on the screen or critical impact parameter, $b_\mathrm{c}$, is well known for any spherically symmetric space-time \citep{perlick2022calculating}; for the RZ metric it reads
\begin{align}
    b_\mathrm{c} = \frac{r_\mathrm{c}}{|N(r_\mathrm{c})|}\quad \text{with} \quad \frac{\mathrm{d}}{\mathrm{d}r}\left(\frac{r^2}{N(r)^2}\right)\Bigg\rvert_{r_\mathrm{c}}=0\;.\tag{24} \label{eq:24}
\end{align}  
From equations \eqref{eq:5}, \eqref{eq:6}, we know that $N=N(r,\epsilon,a_i)$, $i\in \mathbb{N}$, so $r_\mathrm{c}$ and $b_\mathrm{c}$ depend on these same parameters via equation \eqref{eq:24}. In particular, if we fix all the parameters to be zero except for $\epsilon$, $r_\mathrm{c}(\epsilon)$ and its projected radius, $b_\mathrm{c}(\epsilon)$, decrease on the interval $]-1;0.5]$, being $r_\mathrm{c}=3M$ and  $b_\mathrm{c}=3\sqrt{3}M$ for $\epsilon=0$. If only $a_1$ varies, $r_\mathrm{c}(a_1)$ and $b_\mathrm{c}(a_1)$ are decreasing functions too and the Schwarzschild values are retrieved for a vanishing $a_1$. The second and third panels of Figure \hyperref[fig:Parameters]{5} provide a visual representation. We stress that the lensing bands always encompass the critical curve, so that its displacement sets in motion their inner edges.\\
\indent Also the radial position of the ISCO, $r_\mathrm{ms}$, only depends on $N(r,\epsilon,a_i)$, $i\in \mathbb{N}$, via equation \eqref{eq:11}, and it plays a role in the computation of the 4-velocity profile of the emitter, $\boldsymbol{u}^\mathrm{em}$, for a Keplerian disc (see paragraph \hyperref[sec:keplerian]{4.1.1}), influencing then the value of its redshift factor, $g$, via equation \eqref{eq:17}. On the treated intervals, both $r_\mathrm{ms}(\epsilon)$ and $r_\mathrm{ms}(a_1)$, with vanishing $a_1$ or $\epsilon$ respectively, are decreasing functions (see last panel of Figure \hyperref[fig:Parameters]{5}), and  $r_\mathrm{ms}=6M$ for a Schwarzschild black hole with $\epsilon=a_1=0$.\\
\indent To summarise, the values of $r_\mathrm{H}$, $b_\mathrm{c}$, and $r_\mathrm{ms}$ for a RZ non-standard black hole with positive individual parametric deviations, with $\epsilon\leq0.5$, are smaller than in the Schwarzschild case, while they are bigger for $-1<\epsilon<0$, $-1<a_1<0$. This forces direct or lensed intensity's peaks to follow the same tendencies.\\
\indent Finally, we note that all the parameters of the RZ metric influence the null geodesic motion in such a way that the impact locus on the screen of a trajectory reaching a given point around the black hole is modified by them. However, the parameters $b_1$ and $a_0$, $b_0$ (when PPN constraints are satisfied) engender minor modifications on the analysed ray-traced image characteristics, as it is reported in Appendix \hyperref[app:parameters]{C}.

\subsection{Impact of the astrophysics on the observed intensity}
\begin{figure*}[ht!]
\label{fig:redshift}
   \resizebox{\hsize}{!}
    {\includegraphics{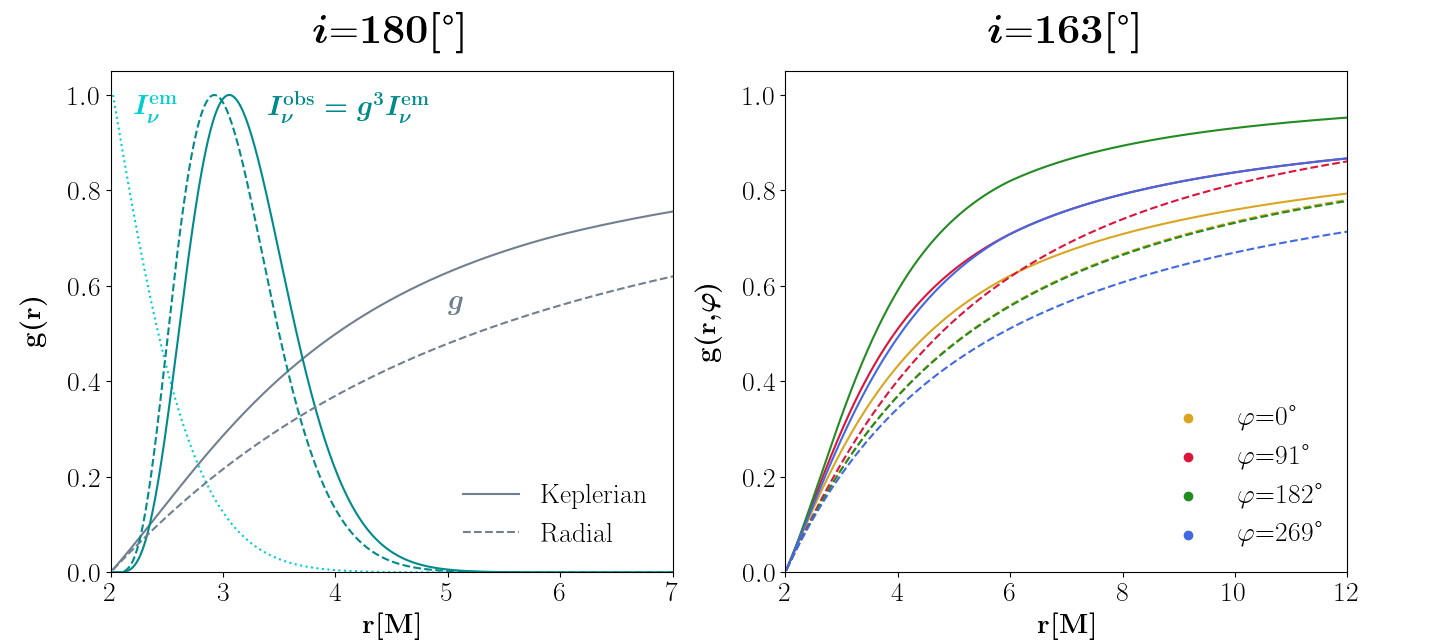}}
    \caption{Redshift profiles in a Schwarzschild space-time for a disc with Keplerian velocity (solid lines) or radially infalling (dashed lines) seen face-on (left panel) or with inclination $i=163^\circ$(right panel). The $r$, $\varphi$ coordinates are those of the emission point of the photon. On the left panel, the emitted intensity with fiducial parameters and $\theta_\mathrm{B}=\mathrm{const}$ (dotted line) and its corresponding redshifted profiles (teal curves) are also traced.}
\end{figure*}
We detail here the effects of various configurations of the disc on the observed intensity, $I^\mathrm{obs}_\nu$, on which the measurements of the image analysis were performed, and on the key quantity needed to compute it, that is, the redshift factor, $g$ (see equations \eqref{eq:16} and \eqref{eq:17}). We recall that when $g>1$ the observed frequency appears to be blueshifted, while if $g<1$, it is redshifted.\\
\indent As mentioned above, at constant $\theta_\mathrm{B}$ all our emission profiles, $I^\mathrm{em}_\nu$, are radially decreasing in the pertinent range, therefore peaking at $r_\mathrm{inner}=r_\mathrm{H}$. As for the observed profile, $I^\mathrm{obs}_\nu$, given that the redshift factor, $g$, always tends asymptotically to $g=1$, starting from $g=0$ at $r_H$, as it is recalled in Appendix \hyperref[app:redshift]{B}, the position of the peak of the observed intensity is moved to a further radius with respect to the one of the event horizon as it is shown in the left panel of Figure \hyperref[fig:redshift]{6}, reaching the screen with a corresponding larger impact parameter (see Figure \hyperref[fig:apparent_radii]{2}). The observed shift is of course related to the viewing angle and the shape of the velocity profile of the disc. As for the width, if the gravitational redshift kills the profile on the left side of the peak, closer to the black hole, its right decay is mainly governed by the emission falloff.\\
\indent If the disc is seen face-on, that is, if $i=0^\circ$, the redshift factor, $g$, is just a function of the radial coordinate, $r$, preserving the axisymmetry of the problem, and depends on the dynamics of the disc (see left panel of Figure \hyperref[fig:redshift]{6}): for every radial distance, $r$, $g_\text{Keplerian}\geq g_\text{Radial}$ and both redshift factors tend to 1 from below (see left panel of Figure \hyperref[fig:redshift]{6}). Besides, photons along a given direction $\psi$ on the screen are emitted from a fixed $\varphi$ on the disc\footnote{In Kerr, a given angle $\psi$ on the observer's screen does not correspond to a specific angle around the black hole \citep{johnson2020universal}.} (see Figure \hyperref[fig:velocity]{3}). This means that the ray-traced intensity would be the same for every direction $\psi$ and, consequently, the intensity peaks are superposed to apparent isoradial curves like the ones illustrated in Figure \hyperref[fig:apparent_radii]{2} and the measured width would be constant along all the polar directions on the screen.\\
\indent This azimuthal symmetry of the observed intensity is lost if the dynamical\footnote{For a static disc, $g$ continues to be independent of $\varphi$, because in that case only the time components of the dot products are non-vanishing.}disc is seen with some inclination and the $g(r,\varphi)$ profile of course still depends on the velocity of the disc (see right panel of Figure \hyperref[fig:redshift]{6}). Put differently, when some inclination is present, the well-known beaming effect, causing a boosting of the apparent luminosity of the part of the source moving towards the observer, emerges. It is visualised in Figure \hyperref[fig:velocity]{3}, the observer being located at $\varphi=270^\circ$: for the radial accretion flow velocity field, the approaching side corresponds to $\varphi\in[0^\circ;180^\circ]$ (upper side of the image) with a maximum at $90^\circ$, whereas for a anticlockwise Keplerian velocity profile, the approaching side coincides with $\varphi\in[90^\circ;270^\circ]$ (left side of the image), with a maximum at $180^\circ$. These variations, comparable to those of $\sin\varphi$ and $-\cos\varphi$ respectively, are coherent with the dependencies in $\varphi$ in the redshift factor when assuming a Minkowski space-time, whose derivation is provided in Appendix \hyperref[app:redshift]{B}.

\begin{figure*}[ht!]
\label{fig:peaksnomag}

   \resizebox{\hsize}{!}         {\includegraphics{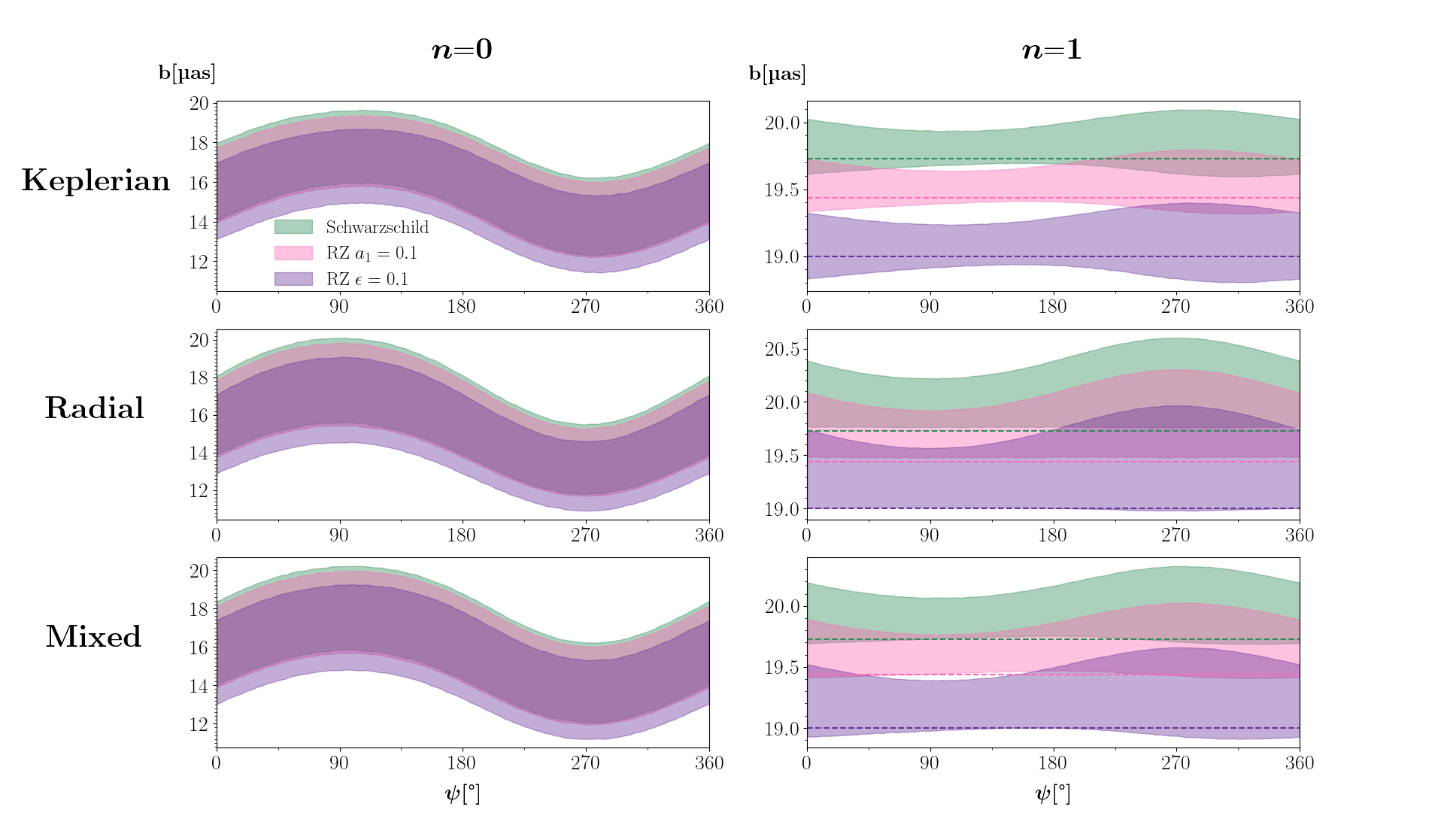}}
    \caption{Positions of the $n=0$ (left panels) and $n=1$ (right panels) peaks, observed at 345 GHz and $i=163^\circ$, for all the possible emission configurations with fixed $\theta_\mathrm{B}=90^\circ$ in the case of a Schwarzschild black hole (green), a RZ black hole with $a_1|_{\,\epsilon=0}=0.1$ (pink), and a RZ black hole with $\epsilon\,|_{\,a_1=0}=0.1$ (violet). Each row corresponds to a different velocity: Keplerian, radial or mixed. Critical curves are represented as dashed lines.}
\end{figure*}

\section{Astrophysical-geometrical degeneracy}
\label{sec:degeneracy} 

If one knew exactly the emission process and the dynamics that govern the disc, deductions on the underlying geometry could be 
undoubtedly made through the electromagnetic signatures of the accretion onto the black hole. Nonetheless, bearing in mind our ignorance of the complex astrophysical mechanisms involved \citep{gralla2021M87}, the question is to know if, in spite of all the astrophysics uncertainties, there are still some observables that permit one to distinguish between a Schwarzschild space-time and a non-standard metric. In the two following paragraphs, we explore the feasibility of the previous statement using the two measured quantities introduced in Section \hyperref[sec:analysis]{6}, namely the width and the peak position of the observed intensity along the polar angles on the observer's screen. We additionally separate the simplistic case in which the angle between the direction of emission and the magnetic field, $\theta_\mathrm{B}$, is fixed to be $90^\circ$ from the one in which it is computed numerically via equation \eqref{eq:18.1}, in order to examine the importance of including the variation in the $\theta_\mathrm{B}$ angle caused by the space-time curvature. The conclusions made are very similar for both the observing frequencies considered in Table \hyperref[tab:simu]{3} (see Appendix \hyperref[app:parameters]{C} for an explanation), and the plots presented correspond to images simulated at $\nu^\mathrm{obs}=345$ GHz. 

\subsection{Observable disentanglement of the peak positions}
\label{sec:disentangle}

\begin{figure*}[ht!]
\label{fig:peaksmag}
   \resizebox{\hsize}{!}         {\includegraphics{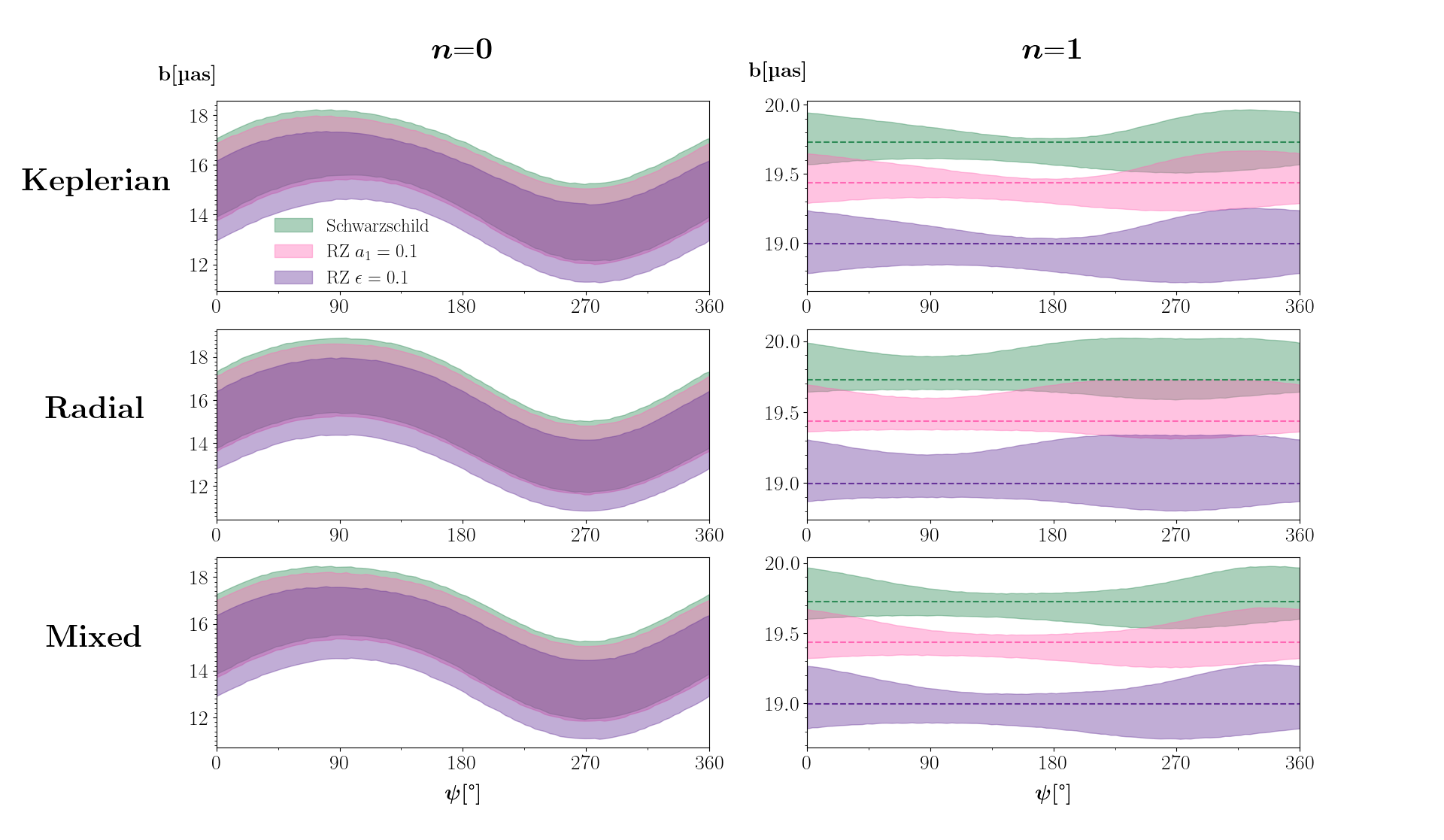}}
    \caption{Positions of the $n=0$ (left panels) and $n=1$ (right panels) peaks, observed at 345 GHz and $i=163^\circ$, for all the possible emission configurations computed including the real value of the synchrotron pitch angle, $\theta_\mathrm{B}$, for a vertical magnetic field in the case of a Schwarzschild black hole (green), a RZ black hole with $a_1|_{\,\epsilon=0}=0.1$ (pink), and a RZ black hole with $\epsilon\,|_{\,a_1=0}=0.1$ (violet). Each row corresponds to a different velocity: Keplerian, radial or mixed. Critical curves are represented as dashed lines.}
\end{figure*}

The direct and first-lensed peaks positions for all possible combinations of the emission parameters of Table \hyperref[tab:astro]{2} were confined in the different bands of Figures \hyperref[fig:peaksnomag]{7} and \hyperref[fig:peaksmag]{8}, each band corresponding to one of the space-times considered in Table \hyperref[tab:BH]{3}: Schwarzschild in green, RZ with $a_1=0.1$ and $\epsilon=0$ in pink, RZ with $\epsilon=0.1$ and $a_1=0$ in violet. The three investigated velocities of the disc, specifically Keplerian, radial, and mixed with a sub-Keplerian predominant part plus radial component, were explored separately in the three rows of the plots.\\ 
\indent Since the three bands are widely superposed in the first columns of Figures \hyperref[fig:peaksnomag]{7} and \hyperref[fig:peaksmag]{8}, we can state that astrophysical uncertainties dominate over small geometric deviations in the determination of the $n=0$ peaks and this regardless of the dynamics of the disc or the inclusion of $\theta_\mathrm{B}$. We remark that this is especially true for the Schwarzschild band, in green, and the $a_1|_{\,\epsilon=0}=0.1$ pink band which almost entirely cover each other because the event horizon, where all the emitted intensity profiles reach their maximum for a constant $\theta_\mathrm{B}$ (see Section \hyperref[sec:analysis]{6}), is located at the same radial distance in these two cases. In other words, the other modifications induced by the non-standard metric, that is, on the geodesic motion, the redshift factor, and the radius of the ISCO (see Section \hyperref[sec:analysis]{6}), are not sufficient to generate a significant observable difference on the $n=0$ peaks position and this is all the more true when varying only $b_1$, as illustrated in Appendix \hyperref[app:parameters]{C}. Even in the case of $\epsilon|_{a_1=0}=0.1$, whose event horizon's radius is $r_\mathrm{H}\simeq 1.81M$, corresponding to a mean projected shift of about $0.84$ µas on the screen with respect to the Schwarzschild horizon, the measured positions of the peaks are highly degenerate between the green and violet bands. Higher parameters deviations lead, unsurprisingly, to more visible divergences, but even for the extreme positive values of equation \eqref{eq:7}, $a_1|_{\,\epsilon=0}=1$ and $\epsilon|_{a_1=0}=0.5$, the astrophysical-geometrical degeneracy is not broken for all the profiles, as shown in the first column of Figure \hyperref[fig:higher]{C.7} of Appendix \hyperref[app:parameters]{C}.\\
\indent On the contrary, even being agnostic of the astrophysical emission, at least in the parameter space spanned in this paper, there is still the possibility to distinguish the strongly lensed observable features produced by different space-time metrics. In other terms, as expected, the geometry of the space-time impacts more dominantly the secondary images of the accretion disc. 
More precisely, for a non-standard black hole, the majority of the emission profiles with fixed $\theta_\mathrm{B}$ of a radially infalling disc or of a disc with mixed velocity produce distinguishable peaks (see bottom right panels of Figure \hyperref[fig:peaksnomag]{7}), and all of them become completely disentangled in the Keplerian case (see top right panel of Figure \hyperref[fig:peaksnomag]{7}). Besides, if the real value of $\theta_\mathrm{B}$ was included, the degeneracy is always broken for all possible motions of the disc, as presented in the second column of Figure \hyperref[fig:peaksmag]{8}. This can be understood by the fact that subjacent space-time plays an additional role in this second case, determining the lensing effects needed in the computation of $\theta_\mathrm{B}$. A detailed explanation of the impact of $\theta_\mathrm{B}$ on the peak positions, also in the case of a radial magnetic field, is provided in Appendix \hyperref[app:theta_mag]{C.3}. Again, as in the $n=0$ case, the distance from the Schwarzschild green band is more pronounced when the position of the event horizon is modified, as in the violet band,\footnote{We note that these violet peaks, contrary to the pink ones, would already be identified as a deviation from Schwarzschild when applying the purely geometric lensing band approach of \citet{cardenas2024lensing}, marginalising then over all conceivable emission profiles.} and the contrasting conclusion vis-à-vis the $n=0$ 
study is mainly due to a modification of another key feature for the image formation, specifically the photon sphere around which $n=1$ light rays swirl before reaching infinity (see Section \hyperref[sec:analysis]{6}). This does not happen for deviations induced by the parameter $b_1$ (see Appendix \hyperref[app:b1]{C.7}). Here too, bigger parameters create more striking disentanglements, as shown in the second column of Figure \hyperref[fig:higher]{C.7} of Appendix \hyperref[app:parameters]{C}. Also, a similar conclusion was reached when looking at higher-order photon rings (see Figure \hyperref[fig:n2]{C.1} in Appendix \hyperref[app:parameters]{C}).\\
\indent As a remark, we note that no curve of Figure \hyperref[fig:apparent_radii]{2} is retrieved exactly because the redshift factor deforms the emitted profiles differently for various values of the azimuthal angle (see Section \hyperref[sec:analysis]{6}) so the measured peaks along the polar angle on the screen are not simply the projections of circular rings as discussed in \hyperref[par:rings]{2.3}. Finally, in Appendices \hyperref[app:rinner]{C.4} and \hyperref[app:inclination]{C.5}, we complement our analysis by presenting the impacts of an inner radius of the disc that does not match the radial position of the event horizon or of a drastically different inclination of the observer on our conclusions.

\subsection{Tangle of the widths}

The width is not an observable that allowed us to disentangle the effects of the astrophysics and the geometry on the electromagnetic images of accretion discs around supermassive black holes. For small deviations they are completely degenerate and even when taking extreme allowed values of the metric parameters, the degeneracy is globally preserved, this for every image order and disc velocity profile (see two last columns of Figure \hyperref[fig:higher]{C.7}). This is not startling in view of the properties already highlighted in Section \hyperref[sec:analysis]{6}. In fact, the width was measured by taking the difference between the two radial distances at which the intensity is half of the maximum: a smaller one, affected by the gravitational redshift, being indeed sensitive to the variation of the metric, and a larger one being mostly prescribed by the falloff of the emission profile. To a certain extent, it is fair to say that the parametric deviations of the space-time are drowned in the diversity of the considered radiation parameters.

\section{Discussion}
\label{sec:end} 

\subsection{Mass uncertainty}
\label{sec:M/D}
\begin{figure*}[ht!]
	\label{fig:mass}
	\resizebox{\hsize}{!}
	{\includegraphics{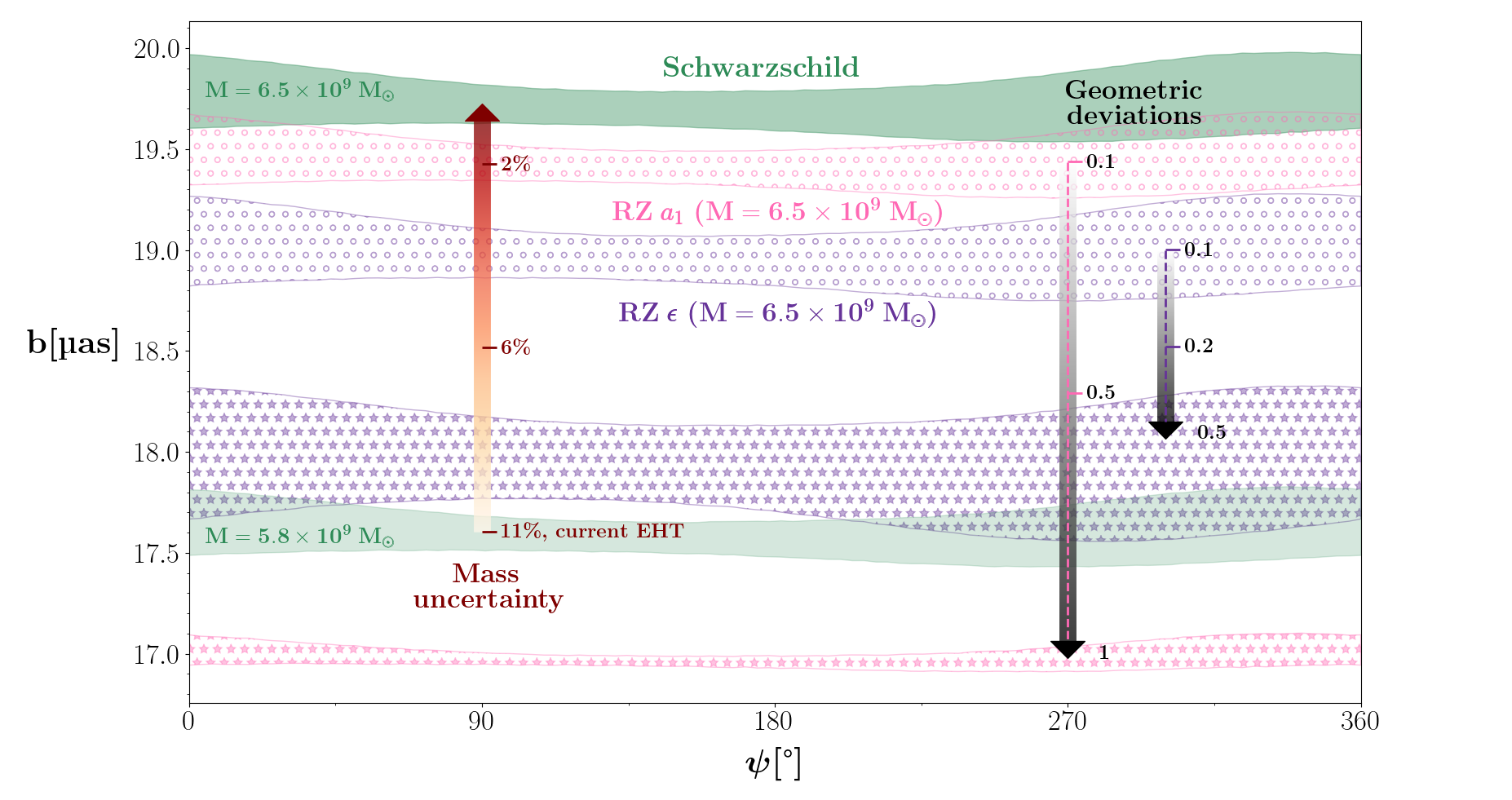}}
	\caption{Positions of the $n=1$ peaks observed at 345 GHz for all the possible emission configurations in the case of a black hole of mass $M=6.5\times10^9$ $\mathrm{M}_{\odot}$ described by a Schwarzschild metric (dark green), a RZ metric with $a_1|_{\,\epsilon=0}=0.1$ or $1$ (dotted pink or starred pink), and a RZ metric with $\epsilon\,|_{\,a_1=0}=0.1$ or $0.5$ (dotted violet or starred violet) and a Schwarzschild black hole of mass $M=5.8\times10^9$ $\mathrm{M}_{\odot}$ (light green). The velocity of the disc is taken to be mixed and the synchrotron pitch angle is computed in the emissivity. The red colour bar shows the error bar from the mean black hole mass value, $M=6.5\times10^9$ $\mathrm{M}_{\odot}$, while the grey colour bars indicate the value of the RZ parameters $a_1$ (pink) and $\epsilon$ (violet).}
\end{figure*}

All the quantities of Figures \hyperref[fig:peaksnomag]{7} and \hyperref[fig:peaksmag]{8} scale with the mass-to-distance ratio, $M/D$, which is expressed in angular units.
From the data of the EHT 2017 campaign, the mass of M87* is measured to be $M=(6.5\pm0.7)\times10^9$ $\mathrm{M}_{\odot}$ under the assumption of general relativity and its distance is adopted to be $D=16.8\pm0.8$ Mpc \citep{EHTVI}. The precision on the mass thus reaches about 11\%, whereas the one on the distance is close to 5\%.
Although the impact of the combined errors on $M$ and $D$ on our results should be examined, we note that the uncertainty on the mass is more than twice as important as that on the distance. Moreover, since the mass is the only parameter defining a Schwarzschild black hole, engendering the structure of its space-time, we assumed for the sake of simplicity that the $D=16.9$ Mpc is perfectly known and that all the uncertainty on $M/D$ is concentrated on the mass. Thus, the question is to know whether a Schwarzschild black hole of lower mass but still inside the EHT uncertainty range could reproduce the $n=1$ characteristic peaks of a non-standard RZ black hole with the mass given in Table \hyperref[tab:BH]{1}. To answer this, we put ourselves in the most optimistic scenario of Section \hyperref[sec:degeneracy]{7}, namely the one in which the disc has a mixed velocity and whose emissivity is given by expression \eqref{eq:21} without further simplifications. Then, we superposed in Figure \hyperref[fig:mass]{9} the bands of first-lensed peaks’ positions in the bottom-right panel of Figure \hyperref[fig:peaksmag]{8} to the ones produced by Schwarzschild black holes of lower masses: $M\in\{5.8, 6.4, 6.49\}$ $\mathrm{M}_{\odot}$ (indicated by red ticks).\\
\indent We observed that the present EHT precision on the value of the black hole mass is not sufficient to discriminate the underlying metric\footnote{The current $M/D$ uncertainty also disables metric tests with photon rings for all alternative spherically symmetric black holes in \citet{wielgus2021photon}.} because small RZ geometric deviations, in dotted bands, can be imitated by a standard black hole with a lower mass contained well inside the uncertainty range. In other words, according to the EHT results, the peaks' band linked to a Schwarzschild space-time can appear everywhere between the green bands. However, improved mass measurements with error bars of the order of $0.1\times10^9$ $\mathrm{M}_{\odot}$, corresponding to 2\% on the red colour bar, could not replicate the violet peaks produced by a RZ metric with $\epsilon|_{a_1=0}=0.1$, and all the small space-time deviations considered in this paper could be detected with a precision on the black hole mass value of the order of $0.01\times10^9$ $\mathrm{M}_{\odot}$. To summarise, small geometric deviations modifying the position of the event horizon could be detected by increasing the precision on the measured mass by about one order of magnitude, whereas one order of magnitude more is needed to disclose a generic small space-time deviation. This conclusion does not change when considering the $n=2$ photon ring, as displayed in Appendix \hyperref[app:parameters]{C}.\\
\indent Conversely, assuming an infinite resolving power, the present EHT precision on the mass is sufficient to isolate the observables linked to the extreme $a_1|_{\epsilon=0}=1$ parametric deviation, and the extreme positive deviation in $\epsilon$, $\epsilon|_{a_1=0}=0.5$, would also become accessible with an uncertainty on $M$ of about $0.4\times10^9$ $\mathrm{M}_{\odot}$, that is, around 6\% on the red colour bar. These results are illustrated by the starred bands in Figure \hyperref[fig:mass]{9}. Thus, there exists a one-to-one relationship between the value of the RZ parameters and the precision on $M/D$ needed to observe their effects. 

\subsection{Relative measurements}
\label{sec:relative}
To go around the stringent precision needed for the $M/D$-dependent measurements of slightly deviating space-times discussed in Section \hyperref[sec:M/D]{8.1}, for an image-plane analysis, one has to consider relative measurements linked, for instance, to the image asymmetry or to respective size of subsequent images. In order to construct such quantities, we introduced the diameter of an $n$-order image, $d^n$, as the distance between two intensity peaks along a polar direction on the screen. We then defined the fractional asymmetry of an $n$-order image, $A^n$, as one minus the ratio between its smallest and biggest diameters and the respective size of successive $n$ and $n+1$ images, $R^{n}$, as the ratio between their mean diameters:
\[A^n\coloneqq1-\frac{d^n_\mathrm{min}}{d^n_\mathrm{max}}\quad\text{and}\quad R^{n}=\frac{d^{n+1}_\mathrm{mean}}{d^n_\mathrm{mean}}\:.\tag{25} \label{eq:25}\]
\indent The asymmetry of the critical curve has already been investigated in \citet{johnson2020universal} and proven not to exceed a few percent for the spinning black hole in M87, being zero percent for a non-rotating central object. We find that under the assumed inclination, the asymmetry is around 1\% and 3\% for the primary image, and at most it reaches 0.3\% for the first photon ring. The latter behaviour is well explained by the convergence of photon rings to the circular critical curve, and it is exacerbated, as expected, for higher-order images, as described in Appendix \hyperref[app:parameters]{C}. We note that both in the $n=0$ and $n=1$ asymmetries, the astrophysical uncertainties are strongly dominating, preventing any test of gravity via this measurement, even for the extreme geometric deviations presented in Appendix \hyperref[app:parameters]{C}. However, if precisions of about 0.2\% could be reached on the asymmetry of the primary image, some constraints could be put on the velocity distribution of the disc, thus reducing the astrophysical assumptions of our models. Also, as the asymmetry of n=1 is highly dependent on the inclination of the source\footnote{And to a lesser extent on spin, as the critical curve \citep{gralla2020lensing}.} (see Appendix \hyperref[app:parameters]{C} for more details), it could be used as a direct measurement of it, complementary to the present estimate where the disc is assumed to be perpendicular to the direction of the jet in M87 \citep{mertens2016kinematics}.\\
\indent We reached similar conclusions for the size ratios of subsequent images. These quantities are highly degenerate between different space-times and even more so for high inclinations. One possibility to try to reduce this degeneracy is to consider a $M/D$-dependent prior on the emission radius \citep{wielgus2020monitoring}.
Although not directly useful to perform tests of general relativity, relative photon-ring sizes could be used to constrain the value of the spin of a rotating black hole \citep{broderick2022measuring} and then reduce again the astrophysical uncertainty.\\ 
\indent To summarise, being able to test RZ deviations from Schwarzschild with relative measurements is unlikely. The main impediment is the fact that under the spherical symmetry assumption, RZ deviations seem to act essentially as scaling modifications. Therefore, we cannot exclude more promising outputs in more exotic space-times with, for example, deformed lensing bands or non-proportional shifts of the event horizon and the photon shell.

\subsection{Conclusion}
In this article, we have revisited the study already undertaken by \citet{bauer2022spherical} and \citet{kocherlakota2022distinguishing} on the prospect to test, via future interferometric observables of black holes such as the ngEHT \citep{ngEHT} or BHEX \citep{BHEX}, theories of gravity within the class of spherically symmetric RZ parametrised space-times \citep{rezzolla2014new}. 
Our main contributions to the effort are the inclusion of physically motivated synchrotron emission profiles in the equatorial plane \citep{vincent2022images}, parametrised according to the results of GRMHD simulations as in \citet{desire2025multifrequency}; the simultaneous analysis of Keplerian, radial, and mixed velocities of the thin accretion disc \citep{pu2016effects}; and the simulations performed at the two higher observing frequencies of the ngEHT and BHEX, 230 and 345 GHz. 
Our focus has been on the case of M87*, but our analysis can be generalised to other supermassive black hole systems.
Our findings are summarised as follows:
\begin{itemize}
	\setlength\itemsep{0.1cm}
      \item All the measured features of the direct image are too polluted by the astrophysical uncertainty on the complex accretion process to be used as a robust probe of the space-time geometry. This is consistent with the results obtained for spherical accretion models \citep{bauer2022spherical,kocherlakota2022distinguishing}. Therefore, the present EHT observations, even if performed at higher observing frequencies, do not have the necessary resolution to detect a violation of general relativity without additional constraints on the properties of the accretion disc.
      Stated differently, the discriminatory test proposed in this paper requires access to at least the $n=1$ photon ring.\\
      \item The width of the $n=1$ intensity profiles along the polar angles on the screen are highly degenerate because the intrinsic emission profile of the disc governs their fall-off \citep{gralla2019black}. In other words, space-time signatures on the width of the observed profiles can be mimicked by variations in the astrophysical configuration, so the width of the $n=1$ photon ring is not a reliable quantity when looking for geometric modifications.\\
      \item The peak positions of the $n=1$ intensity profiles along the polar angles on the screen are the only investigated features that allow the deviations induced by the space-time to be disentangled from those produced by a class of viable emission profiles,\footnote{Within general relativity, the $n=1$ photon-ring radii are mainly influenced by the Kerr black hole spin, rather than by astrophysical diversity \citep{desire2025multifrequency}. Besides, our assertion holds for some, but not all, alternative spherically symmetric black holes in \citet{eichhorn2023universal}, $n=2$ being otherwise needed.} in agreement with the conclusions of preliminary works \citep{kocherlakota2022distinguishing}. 
      Also, the degeneracy is broken unambiguously if the mass over distance is well constrained and for all possible motions of the disc, ranging from a Keplerian to a radially infalling velocity, when the synchrotron emissivity is computed including the value of the synchrotron pitch angle, $\theta_\mathrm{B}$. This is true for both components of a poloidal magnetic field. Since the velocity of realistic accreted matter is expected to have a predominant Keplerian fraction, this result is promising, although challenging, for M87* space-time's constraints via future very-long-baseline interferometry (VLBI) missions.\\
      \item Constraints on the black hole mass-to-distance estimate that are significantly more stringent than the present EHT ones, passing from an uncertainty on $M/D$ of about 11\% \citep{EHTVI} to at least 2\%, are needed to perform the discriminatory test of general relativity proposed in this paper using solely the intensity peak position of the first, or second, photon ring for small geometric deviations. On the other hand, some extreme metric deviations are already accessible with the current precision on $M/D$ given a sufficient resolving power.\\
      \item Relative measurements, independent of $M/D$, such as image asymmetry or images size ratio, cannot be used directly to distinguish a spherically symmetric conservative space-time.\footnote{Relative measurements are sufficient for non-parametrised alternatives to the Schwarzschild metric \citep{wielgus2021photon,eichhorn2023universal}.} Yet, they are complementary probes of the astrophysical set-up, which is at present the main impediment to the identification of small observational deviations from general relativity.  \end{itemize}

\subsection{Limitations of this work and perspectives} A series of approximations are present in our modelling, and the results presented only stand under such simplifying hypotheses.\\
\indent First of all, our equatorial disc was taken to be optically and geometrically thin so that absorption effects along the geodesic motion are neglected, and our simulations did not include any turbulence or observing noise. However, the luminosity of M87*, about $3.6\times10^{-6}L_\mathrm{Edd}$ \citep{prieto2016central}, where $L_\mathrm{Edd}$ is the Eddington luminosity, is very low, corresponding to a radiatively inefficient accretion flow \citep{narayan1995explaining}, and RIAFs nuclei are expected to be surrounded by a hot geometrically thick disc \citep{yuan2014hot}.
The integration of the above mentioned elements could, on one hand, weaken and spoil the lensed features of the image and, on the other hand, add astrophysical uncertainty and thus degeneracy. Nevertheless, EHT data favour the scenario in which the emission is dominated by that of the equatorial plane, and at 345 GHz absorption effects are reasonably weak \citep{EHTV}, so the framework of this paper remains adapted to model future VLBI observations. In any case, it is the objective of future work to expand the query of this paper to more realistic geometrically thick discs, compute the full radiative transfer \citep{vincent2022images}, and incorporate statistical fluctuations \citep{lee2021disks}.\\
\indent Secondly, here we only varied one parameter at a time, fixing the others at their Schwarzschild value. Nonetheless, particular parametric combinations, taken of course into their theoretically allowed ranges \citep{kocherlakota2022distinguishing}, may reduce the deviations induced by the metric, then increasing the entanglement with the astrophysical possibilities. If it would be worth analysing the outputs of such choices, it would be even more compelling to create a new framework in which specific parametric deviations have clear physical interpretations.\\
\indent Third, the discussions made here concern image-plane time-averaged images, but it would be of great interest to know if the obtained results could be translated in terms of interferometric signals and be exploited for a direct real-data analysis (see for example \citet{cardenas2023prediction}).\\
\indent Finally, we only considered spherically symmetric space-times, but realistic black holes are expected to be rotating. Since the spin of M87* is not well constrained, this would represent an additional astrophysical parameter to scan, and the spin may mimic the observational divergences produced by metric deviations. Nevertheless, the relative measurements presented in Section \hyperref[sec:relative]{8.2} could be used to measure its value \citep{johnson2020universal,broderick2022measuring}. A generalisation of the RZ metric to circular axisymmetric space-times already exists \citep{konoplya2016general}, and its implementation in a ray-tracing code could be straightforward.

\begin{acknowledgements}
      I.U. and F.H.V. warmly thank Andrew Chael for insightful comments and constructive suggestions. M.W. is supported by a Ramón y Cajal grant RYC2023-042988-I from the Spanish Ministry of Science and Innovation.

\end{acknowledgements}

\bibliographystyle{aa}
\bibliography{Bibliography.bib}

\begin{appendix} 
\section{Inclination effects on projections}
\label{app:lensing_bands}

\begin{figure}[ht!]
   \centering
   \includegraphics[width=\hsize]{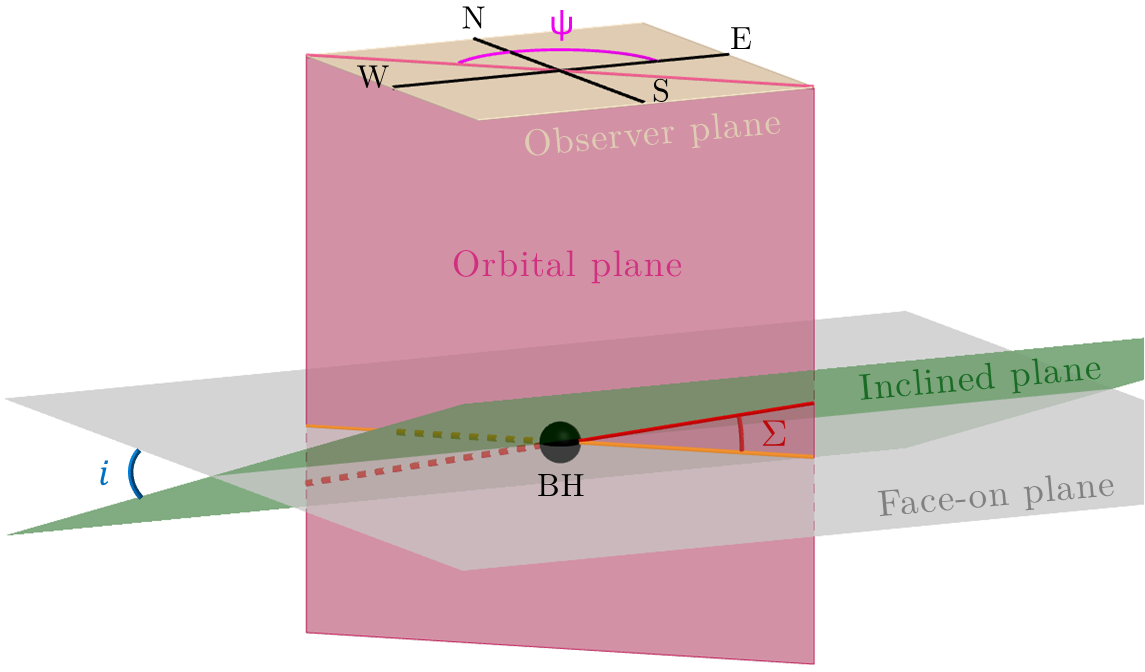}
      \caption{Intersections between a face-on (grey) or inclined (green) plane, the yellow screen of a distant observer, and a pink orbital plane.}
\label{fig:3D}
\end{figure}

To understand lensing band deformations caused by the inclination between an equatorial plane and a distant observer, it is useful to employ the bending number, $w$.\\
\indent We considered the set-up of Figure \hyperref[fig:3D]{A.1}. The far-away observer's yellow screen, labelled by polar coordinates, $(b,\psi)$, as in Figure \hyperref[fig:theoretical_features]{1}, is situated at $\varphi=\varphi_\infty=0\bmod 2\pi$ rad. We defined a $\psi$-direction on the observer's screen as a line passing through its centre with a slope defined by the polar angle, $\psi$. From the same arguments exposed in Section \hyperref[sec:img]{2}, it is immediate to see that photons reaching the same $\psi$-direction on the screen, belong to the same orbital plane. We also considered an infinitely large green plane cutting the black hole into two hemispheres and whose approaching side to the observer is directed towards the south of their screen.\footnote{Different orientations of the plane can be easily deduced by a corresponding rotation of the north of the screen.} The approaching side of the plane, rotated around the east-west axis, is inclined at an angle of $i<\pi/2$ rad\footnote{If the equatorial plane contained a disc, the latter is not seen edge-on.} from the face-on grey plane. We called $\Sigma$ the red angle between the intersections, with the orbital plane, of the inclined plane and of the face-on plane. For the east-west $\psi=0$ direction on the screen, $\Sigma$ is zero because the pink plane passes precisely along the intersection of the green and grey ones. For the north-south $\psi=\frac{\pi}{2}$ direction, $\Sigma$ is equal to $i$, the relative inclination between the green and grey planes. When rotating the orbital plane from east to north, $\Sigma$ takes intermediate values; the orbital plane of Figure \hyperref[fig:theoretical_features]{A.1} corresponds indeed to an intermediate $\psi$-direction, for which $0<\psi<\pi$.

\begin{figure}[ht!]
   \centering
   \includegraphics[width=\hsize]{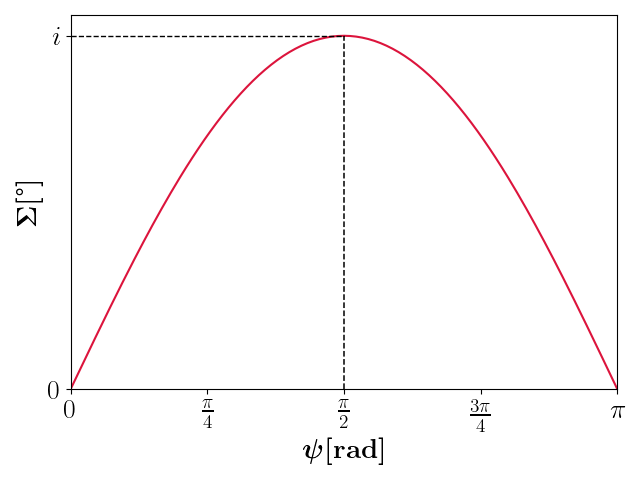}
      \caption{Evolution of $\Sigma$ with the direction on the observer's screen.}
\label{fig:Sigma}
\end{figure}

More precisely, the angle $\Sigma$ varies with the direction encountered by photons on the observer's screen as follows: 
\begin{align*}
\begin{cases}
\:\Sigma = 0 & \text{for } \psi=0\\
\:\Sigma = i & \text{for } \psi=\cfrac{\pi}{2}\\
\:\Sigma = \arccos\left\{\cfrac{\sqrt{1+\tan\psi^{-2}}}{\sqrt{1+\tan\psi^{-2}\,+\tan^2 i}}\right\}&\text{otherwise}\:.
\tag{A1} \label{eq:A1}
\end{cases}
\end{align*}
So, $\Sigma$ increases symmetrically from $\psi=0$ to $\psi=\pi/2$ and from $\psi=\pi$ to $\psi=\pi/2$, as shown in Figure \hyperref[fig:Sigma]{A.2}.\\
\begin{figure*}[ht!]
   \resizebox{\hsize}{!}
            {\includegraphics{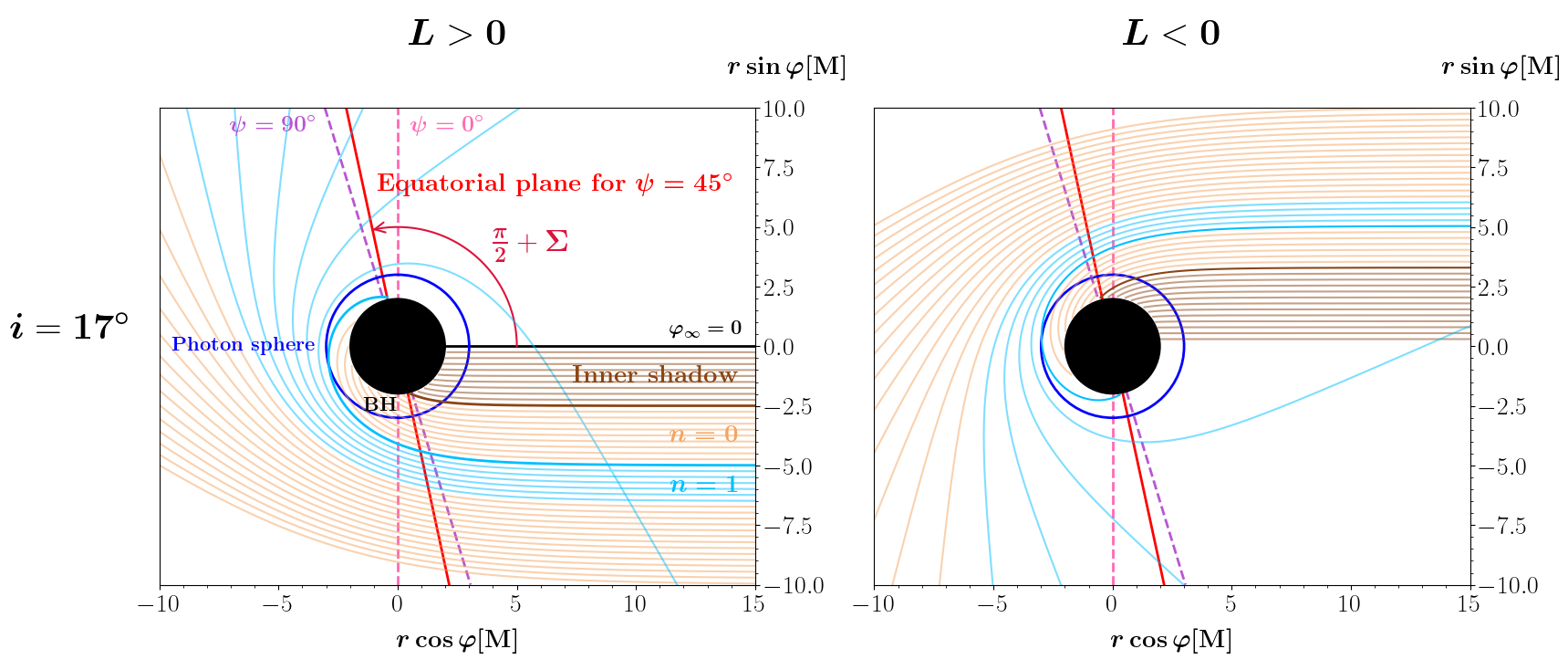}}
    \caption{Null geodesics in a Schwarzschild space-time reaching an observer on the far right. According to their winding number, geodesics can make up the inner shadow (brown), the $n=0$ image (orange), or the $n=1$ photon ring (light blue). We represent the projections of the black hole (black), the photon sphere (blue), and the equatorial plane for different polar directions on the screen: $\psi=0^\circ$ (pink), $45^\circ$ (red), and $90^\circ$ (violet).}
\label{fig:Deflection}
\end{figure*}
We defined the total change, $\Delta\varphi$, of the orbital plane
angle, $\varphi$, and the bending number, $w$, as in \citet{gralla2019black}:
\begin{align*}
	\Delta\varphi\coloneqq\varphi_{\infty}-\varphi_{-\infty}\:,\:\:\: \varphi_{\pm\infty}=\lim_{\hat{s}\to\pm\infty} \varphi\quad\text{and}\quad w\coloneqq\Delta\varphi/2\pi\:.\tag{A2} \label{eq:A2}    
\end{align*}
Calling $\varsigma_\mathrm{\lambda}$ the sign of $\lambda$, we remark that $w=0$ for radial trajectories, $w\to\varsigma_\mathrm{\lambda}1/2$ when $b\to\infty$, while $w\to\varsigma_\mathrm{\lambda}\infty$ for $b\to \tilde{b}$. Having defined the bending number $w$ in equation \eqref{eq:A2}, we propose hereinbelow a classification of lensing bands that generalises previous conventions \citep{gralla2019black, chael2021observing} to arbitrary plane inclinations and polar directions on the observer's screen. With $\varsigma_\mathrm{\lambda}$ the sign of $\lambda$ and $\sigma\coloneqq\Sigma/(2\pi)$, light rays are classified according to their order, $n$, defined as follows:
\begin{align*}
\begin{cases}
\:n=0 & \:\text{if } \: 0 \leq \varsigma_\mathrm{\lambda}w < \cfrac{3}{4}-\sigma  \\*[0.5em]
\,\dfrac{3}{4}-\sigma+\cfrac{n-1}{2} \leq \varsigma_\mathrm{\lambda}w < \cfrac{3}{4}-\sigma +\cfrac{n}{2} & \:\text{if } \: \varsigma_\mathrm{\lambda}w \geq \cfrac{3}{4}-\sigma \:,
\end{cases}\tag{A3} \label{eq:A3}
\end{align*}
Thus, for $n\geq1$, the $n$-th lensing band is the area on the plane of sky including all light rays whose bending number, $w$, is inside the range settled by the order $n$, crossing then the equatorial plane exactly $n+1$ times along their way from the black hole to the observer or vice-versa (see Figure \hyperref[fig:Deflection]{A.3}). Given this definition, the periodicity of $\Sigma$, and so of $\sigma$, leads to the reflection symmetry of the lensing bands with respect to the north-south vertical direction on the observer's screen. Put differently, if the projections of the equatorial plane in Figure \hyperref[fig:Deflection]{A.3} tilt to the left from $\psi=0^\circ$ to $\psi=90^\circ$, they symmetrically come back to the initial perpendicular direction going towards $\psi=180^\circ$, imposing then the same geodesic orders.

\section{Redshift factor}
\label{app:redshift}

\subsection{Asymptotes}
Here, we discuss the limits of $g$ at infinity and at the horizon.\\
\indent At infinite distance from the black hole, it is legitimate to consider that the emitter is static and that the space-time is flat, in other words $\boldsymbol{u}^\mathrm{em}=\boldsymbol{u}^\mathrm{obs}$ and $\boldsymbol{p}^\mathrm{em}=\boldsymbol{p}^\mathrm{obs}$. By means of equation \eqref{eq:17}, the redshift factor at radial infinity is
\[\lim_{r\to\pm\infty} g=1\:.\tag{B1} \label{eq:B1}\]
\indent If the emitter is asymptotically close to the radius of the event horizon, $r_\mathrm{H}$, the radial component of its 4-velocity, $u^r_\mathrm{em}$, cannot be zero because below the radius of the ISCO, $r_\mathrm{ms}>r_\mathrm{H}$, no stable circular orbit is allowed, and the 4-momentum of the emitted photon has to possess a radial component, $p^r_\mathrm{em}$, to escape the gravitational attraction of the black hole and reach the observer at infinity. Besides, the 4-velocity of the emitter is proportional to its 4-momentum and 4-momenta's time and azimuthal components are related to the constants of motion via equations \eqref{eq:1}. Then, at the horizon, the non-vanishing metric components of the RZ metric \eqref{eq:4}, \eqref{eq:5}, \eqref{eq:6}, with $x=1-r_\mathrm{H}/r$, read
\begin{align*}
    \lim_{r\to r_\mathrm{H}} g_{tt}&\coloneqq \lim_{r\to r_\mathrm{H}}-N^2(r)=0\:,
    \tag{B2.1} \label{eq:B2.1}\\
    \lim_{r\to r_\mathrm{H}} g_{rr}&\coloneqq \lim_{r\to r_\mathrm{H}}B^2(r)/N^2(r)=\infty\:,
    \tag{B2.2} \label{eq:B2.2}\\ 
    \lim_{r\to r_\mathrm{H}} g_{\theta\theta}&\coloneqq \lim_{r\to r_\mathrm{H}}r^2=r_\mathrm{H}^2\:,
    \tag{B2.3} \label{eq:B2.3}\\
    \lim_{r\to r_\mathrm{H}} g_{\varphi\varphi}&\coloneqq \lim_{r\to r_\mathrm{H}}r^2\sin^2\theta=r_\mathrm{H}^2\sin^2\theta\:.
    \tag{B2.4} \label{eq:B2.4}
\end{align*}
Thus, the scalar product $\boldsymbol{p}^\mathrm{em}\cdot\boldsymbol{u}^\mathrm{em}=p^\mathrm{em}_{\,\mu}\,u_\mathrm{em}^{\,\mu}$, appearing at the denominator of equation \eqref{eq:17}, diverges because the constants of motion are finite and one can show that so are $p_r^\mathrm{em}$ and $u^r_\mathrm{em}$, by considering, respectively, the normalisation $p^r_\mathrm{em}p_r^\mathrm{em}=0$ and the photon's geodesic equation along $\theta$ or the expression \eqref{eq:12} of a fully radial infall. Thus, when $r$ tends to the radial position of the event horizon, for every equatorial velocity profile of the emitter, $\boldsymbol{u}^\mathrm{em}$, the limit of the redshift factor is
\[\lim_{r\to r_\mathrm{H}} g=0\:.\tag{B3} \label{eq:B3}\]

\subsection{Minkowski approximation}

The Minkowski metric line element in spherical coordinates is
\begin{align*}
    ds^2=-dt^2+dr^2+r^2(d\theta^2+\sin^2\theta d\varphi^2)\:.\tag{B4} \label{eq:B4}
\end{align*}
We recall that the redshift factor, $g$, is given by \eqref{eq:17}
\begin{align*}
  g=\frac{\boldsymbol{p}^\mathrm{obs}\cdot \boldsymbol{u}^\mathrm{obs}}{\boldsymbol{p}^\mathrm{em}\cdot \boldsymbol{u}^\mathrm{em}}\:.\tag{B5} \label{eq:B5}  
\end{align*}
In Minkowski space-time, the 4-momentum of the photon has a simple, geodesic invariant, analytical form \citep{vincent2024polarized}:  
\begin{align*}
   \boldsymbol{p}^\mathrm{obs}=\boldsymbol{p}^\mathrm{em}=\mathbf{e}_t-\sin\iota\sin\varphi\,\mathbf{e}_r+\cos\iota\,\mathbf{e}_\theta-\sin\iota\cos\varphi\, \mathbf{e}_\varphi\:,\tag{B6} \label{eq:B6}
\end{align*}
where $\iota=\pi-i\in[0^\circ;90^\circ]$ for $i\in[90^\circ;180^\circ]$ and where the natural basis vectors, $\boldsymbol{\partial}_\mu$, have been replaced by
\begin{align*}
    \mathbf{e}_t=\frac{\boldsymbol{\partial}_t}{\sqrt{-g_{tt}}},\quad\mathbf{e}_r=\frac{\boldsymbol{\partial}_r}{\sqrt{g_{rr}}},\quad\mathbf{e}_\theta=\frac{\boldsymbol{\partial}_\theta}{\sqrt{g_{\theta\theta}}},\quad\mathbf{e}_\varphi=\frac{\boldsymbol{\partial}_\varphi}{\sqrt{g_{\varphi\varphi}}}\:.\tag{B7} \label{eq:B7}
\end{align*}
As the velocity of a static observer is directly given by $\boldsymbol{u}^\mathrm{obs}=\mathbf{e_t}$, one has $\boldsymbol{p}^\mathrm{obs}\cdot \boldsymbol{u}^\mathrm{obs}=-1$ at the numerator of equation \eqref{eq:B5}.
As for the scalar product appearing at the denominator of $g$ in equation \eqref{eq:B5}, it varies according to the 4-velocity of the emitter. The general expressions of a Keplerian and a radial 4-velocities are
\begin{gather*}
    \begin{aligned}
        &\boldsymbol{u}_\text{Keplerian}=u^t_\text{Keplerian}(\boldsymbol{\partial}_t+\Omega\boldsymbol{\partial}_\varphi)\:,\quad &&\Omega=u^\varphi/u^t=d\varphi/dt\\
        &\boldsymbol{u}_\text{Radial}=u^t_\text{Radial}(\boldsymbol{\partial}_t+\chi\boldsymbol{\partial}_r)\:,\quad\ &&\:\chi=u^r/u^t=dr/dt\:.
    \end{aligned}\tag{B8} \label{eq:B8}
\end{gather*}
The expressions of $\Omega$ and $\chi$ depend on the space-time metric and $u^t$ can be deduced by the normalisation $\boldsymbol{u}\cdot\boldsymbol{u}=-1$. If one takes the expressions of $\Omega$ and $\chi$ for the Schwarzschild space-time, which coincide with the Newtonian results, one gets
\begin{gather*}
    \begin{aligned}
        \Omega&=\frac{1}{r}\sqrt{\frac{M}{r}}&&\Rightarrow\quad u^t_\text{Keplerian}=\sqrt{\frac{r}{r-M}}\\
        \chi&=-\sqrt{\frac{2M}{r}}&&\Rightarrow\quad u^t_\text{Radial}=\sqrt{\frac{r}{r-2M}}\:.
    \end{aligned}\tag{B9} \label{eq:B9}
\end{gather*}
Thus, in the orthonormal basis $\mathbf{e}_\mu$, the 4-velocities read
\begin{gather*}
    \begin{aligned}
        &\boldsymbol{u}_\text{Keplerian}=\sqrt{\frac{r}{r-M}}\left(\mathbf{e}_t+\sqrt{\frac{M}{r}}\mathbf{e}_\varphi\right)\\
         &\boldsymbol{u}_\text{Radial}=\sqrt{\frac{r}{r-2M}}\left(\mathbf{e}_t-\sqrt{\frac{2M}{r}}\mathbf{e}_r \right)\:.
    \end{aligned}\tag{B10} \label{eq:B10}
\end{gather*}
With the previous derivations, it is then immediate to obtain the expressions of the denominator of $g$:
\begin{gather*}
    \begin{aligned}
        &(\boldsymbol{p}^\mathrm{em}\cdot \boldsymbol{u}^\mathrm{em})_\text{Keplerian}=-\sqrt{\frac{r}{r-M}}\left(1+\sqrt{\frac{M}{r}}\sin\iota\cos\varphi \right)\\
         &(\boldsymbol{p}^\mathrm{em}\cdot \boldsymbol{u}^\mathrm{em})_\text{Radial}=-\sqrt{\frac{r}{r-2M}}\left(1-\sqrt{\frac{2M}{r}}\sin\iota\sin\varphi \right)
    \end{aligned}\tag{B11} \label{eq:B11}
\end{gather*}
and the analytical formulae of the redshift are thus
\begin{gather*}
\begin{aligned}
    g^\text{Minkowski}_\text{Keplerian}&=\left[\sqrt{\frac{r}{r-M}}\left(1+\sqrt{\frac{M}{r}}\sin\iota\cos\varphi \right)\right]^{-1}\\
    g^\text{Minkowski}_\text{Radial}&=\left[\sqrt{\frac{r}{r-2M}}\left(1-\sqrt{\frac{2M}{r}}\sin\iota\sin\varphi \right)\right]^{-1}\:.
\end{aligned}\tag{B12} \label{eq:B12}
\end{gather*}
We note that for a disc seen face-on from below, $i=\pi\Rightarrow \iota=0^\circ$ and the Minkowski redshift factors are independent of $\varphi$. 

\section{Additional parametric explorations}
\label{app:parameters}

The aim of this section is to justify the choice of the relevant parameters considered in the main text and to survey a range of additional ones which go beyond the scope of this paper. All the simulations of this Appendix, unless differently stated, take into account the synchrotron pitch angle in the computation of the emissivity \eqref{eq:21} and are performed at an observing frequency of 345 GHz. Also, we generally only consider the peaks position of the first photon ring of a disc with a mixed velocity profile around a black hole of mass $M=6.5\times10^9$ $\mathrm{M}_{\odot}$ surrounded by a vertical magnetic field.

\subsection{Observing frequency}

The results of Figures \hyperref[fig:peaksnomag]{7} and \hyperref[fig:peaksmag]{8}, computed for $\nu_\mathrm{em}=345$ GHz, as well as the corresponding interpretations are very similar to the ones obtained at 230 GHz, so it is of no interest to plot or repeat them in the case of the higher observing frequency, and we therefore only briefly explain why this resemblance is unsurprising. The observing frequency only intervenes in the computation of the observed intensity in equation \eqref{eq:16}, the emitted intensity and frequency being unchanged. Thus, the value of the redshift factor of equation \eqref{eq:17} is modified, but not its radial dependence, so that the main impact of the observing frequency is on the value of radiative flux density received in a given band and not on the peak position of normalised intensity cuts.\\
\indent Beyond gravity tests, we note the importance of observing at several frequencies to get physical insights on the complex and multi-wavelength emission process and to improve the interferometric coverage for image reconstruction \citep{ngEHT}.

\subsection{Higher-order photon rings}

\begin{figure}[ht!]
	\label{fig:n2}
	\centering
	\includegraphics[width=\hsize]{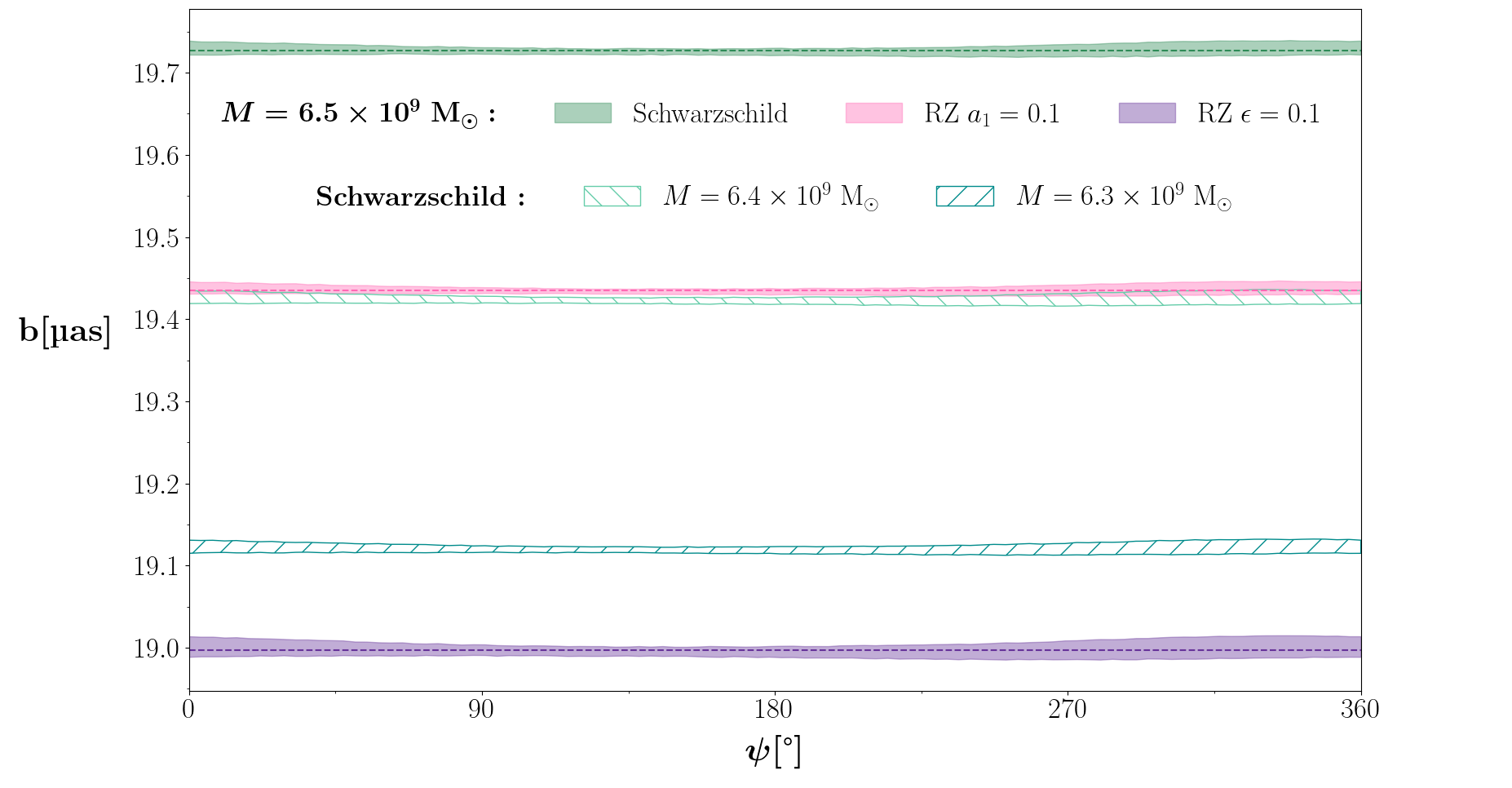}
	\caption{Positions of the $n=2$ peaks for all the possible emission configurations in the case of black holes of mass $M=6.5\times10^9$ $\mathrm{M}_{\odot}$ described by
	a Schwarzschild metric (green), a RZ metric with $a_1|_{\,\epsilon=0}=0.1$ (pink), a RZ metric with $\epsilon\,|_{\,a_1=0}=0.5$ (violet) and Schwarzschild black holes of mass $M=6.4\times10^9$ $\mathrm{M}_{\odot}$ (hatched turquoise) and $M=6.3\times10^9$ $\mathrm{M}_{\odot}$ (hatched teal). Dashed lines represent critical curves.}
\end{figure}

We simulated the second photon ring adaptively in the $n=2$ lensing band, with radial resolution $\mathrm{res}^{b,\:n=2}=0.001$ µas.\\
\indent This higher lensed observable appears to have more clearly separated intensity peaks positions given a sufficient resolving power. Nonetheless, as for the $n=1$ image, these peaks cannot be used as a discriminating probe of any conservative space-time when the uncertainty on the mass-to-distance estimate is too high; here a precision of 3\% is needed. This limited relaxation of the $M/D$ precision constraint can be understood by recalling that photon rings track the critical curve for increasing $n$ and that the position of the critical curve is a $M/D$ highly dependent quantity.\\
\indent The relative measurements introduced in Section \hyperref[sec:relative]{8.2} are again very degenerate even for this more strongly lensed observable, the asymmetry being of the order of 0.01\% for all considered space-times. 

\subsection{Magnetic field configuration}
\label{app:theta_mag}

As already mentioned in the main text, for a MAD system the presence of a strong poloidal magnetic field is favoured \citep{narayan2003magnetically,EHTVIII}. Thus, it is of interest to consider the radial component of the magnetic field in addition to the vertical one. In principle, the advection of the rotating fluid may also create a magnetic-field azimuthal component \citep{yuan2014hot}, but the latter is not investigated in this paper.

\begin{figure}[ht!]
	\centering
	\includegraphics[width=\hsize]{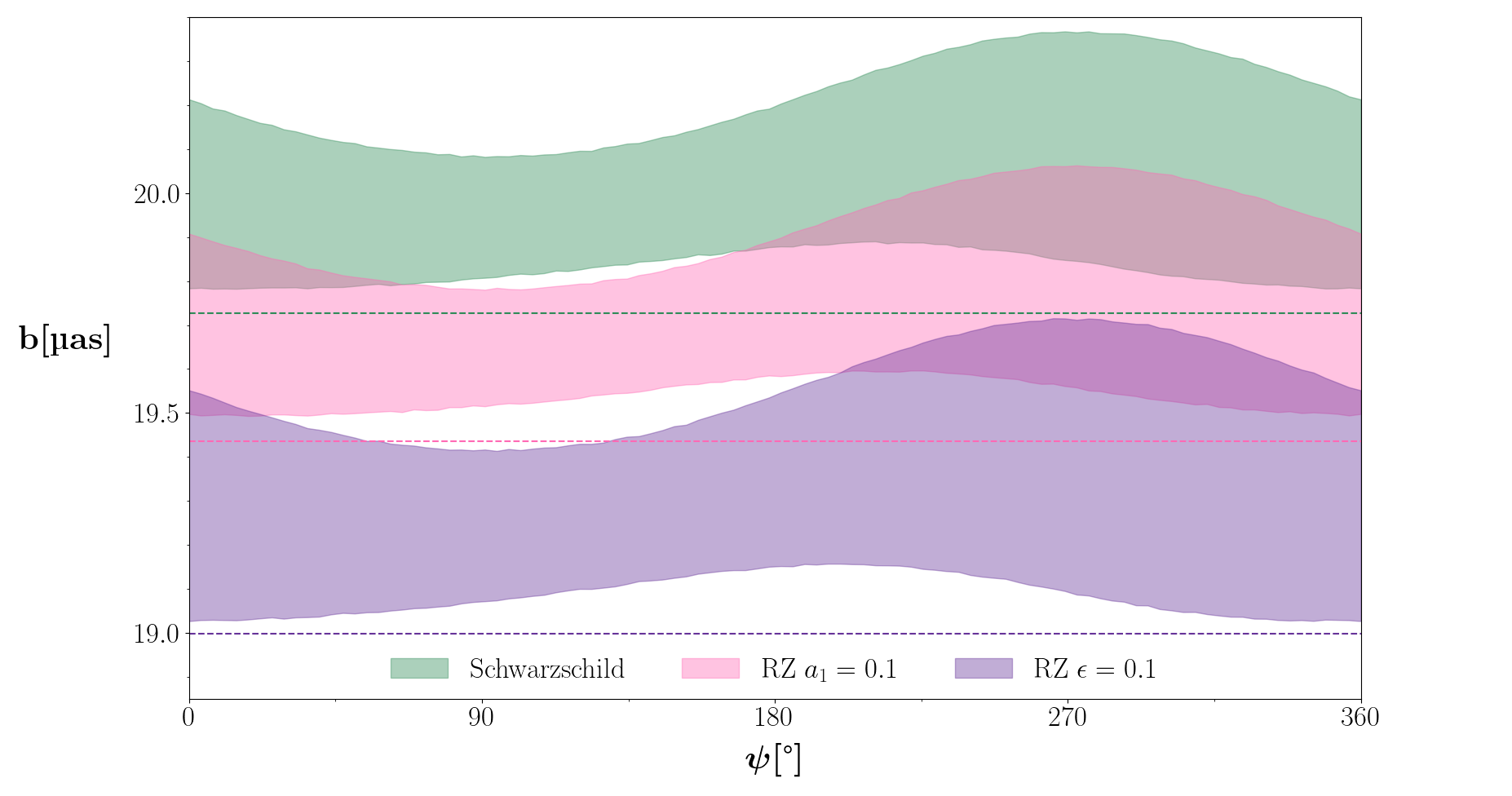}
	\caption{Positions of the $n=1$ peaks for all the possible emission configurations in the case of a Schwarzschild black hole (green), a RZ black hole with $a_1|_{\,\epsilon=0}=0.1$ (pink), and a RZ black hole with $\epsilon\,|_{\,a_1=0}=0.1$ (violet). Critical curves are represented as dashed lines and the magnetic field is radial.}
	\label{fig:RadialB}
\end{figure}

\begin{figure*}[ht!]
	\label{fig:thetaB}
	\resizebox{\hsize}{!}         {\includegraphics{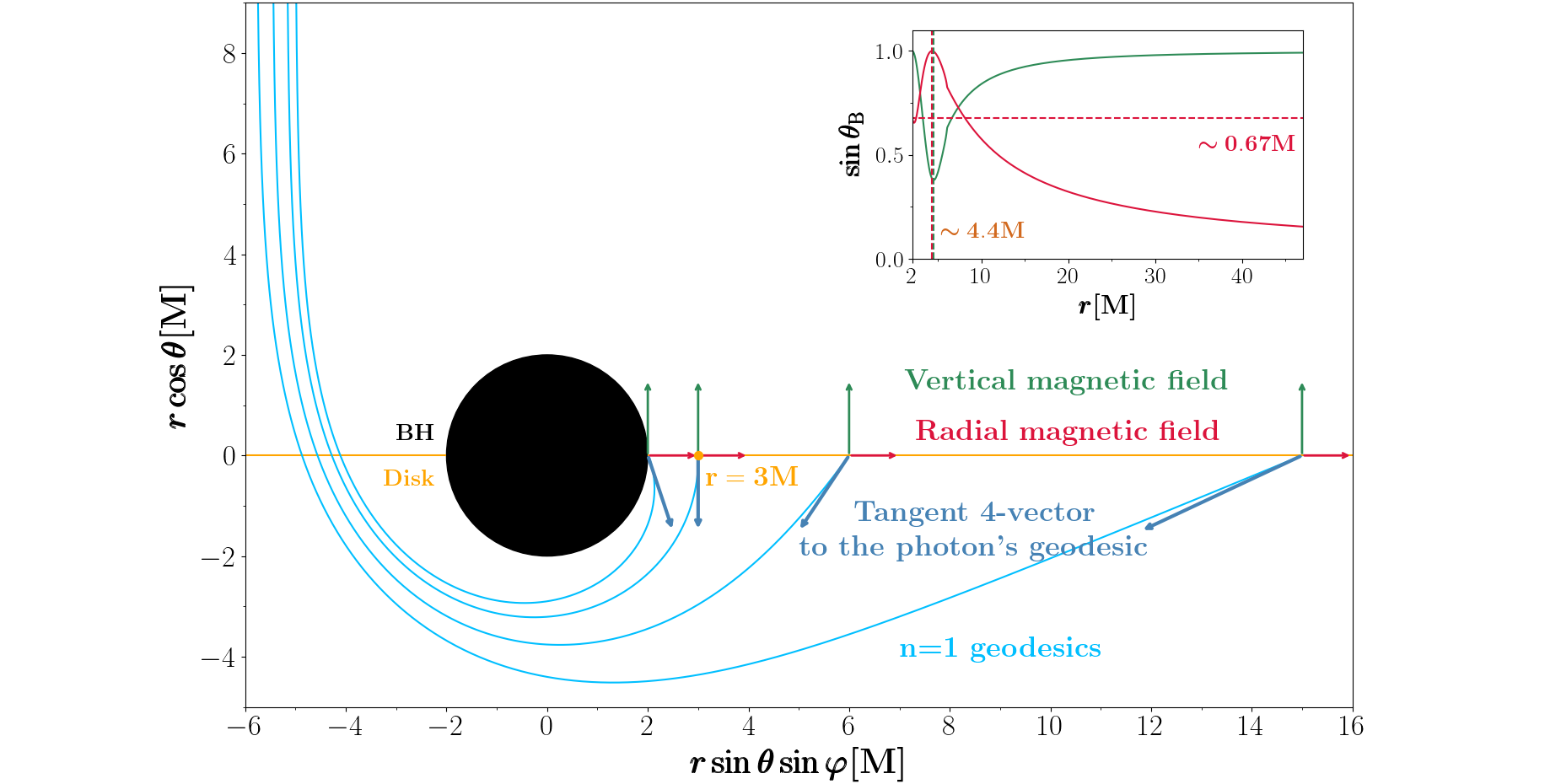}}
	\caption{$n=1$ geodesics emitted at $r=2,3,6,15\,M$ and directions of emission (blue) as well as those of a vertical (green) and a radial (red) magnetic field. The inset shows the value of the pitch angle, $\theta_\mathrm{B}$, of the first lensed photons for the two, vertical and radial, magnetic field configurations in the case of a disc with mixed velocity profile.}
\end{figure*}

The peak positions obtained for a radial magnetic field (see Figure \hyperref[fig:RadialB]{C.2}) are equivalent in terms of astrophysical-geometrical degeneracy with respect to the ones assuming a vertical magnetic field in Figure \hyperref[fig:mass]{9}. However, contrary to the vertical case, all bands in Figure \hyperref[fig:RadialB]{C.2} are located strictly above the critical curve. We analysed the reasons of this different behaviour.\\
\indent First of all, it is worth noticing that the synchrotron pitch angle, $\theta_\mathrm{B}$, only appears in the emissivity of equation \eqref{eq:21} which is then the only quantity affected by a change of the magnetic field configuration. Also, from its definition, that is, the angle between the magnetic field vector and the emission direction in the rest frame of the emitter, it is clear that its value would be different for a same point source if one considers its direct or secondary image, because the directions of emission of a $n=0$ or $n=1$ photon differ. We then treated $n=0$ and $n=1$ separately and considered a disc seen face-on around a Schwarzschild black hole for simplicity.\footnote{We expect our statements to be just slightly modified when adding a small inclination or small metric deviations.}\\
\indent Let $\boldsymbol{k}=\boldsymbol{p}^\mathrm{em}/\hbar$, with $\hbar$ the reduced Planck constant, be the 4-vector tangent to the photon's
geodesic at emission. Its space-like projection, $\boldsymbol{K}$, orthogonal to the 4-velocity of the emitter, $\boldsymbol{u}$, that is, in the rest-frame of the emitter, reads \citep{vincent2024polarized} $\boldsymbol{K} = \boldsymbol{k}+(\boldsymbol{k}\cdot\boldsymbol{u})\,\boldsymbol{u}$.
In the rest frame of the emitter, the magnetic field 4-vector is also orthogonal to the 4-velocity of the emitter: $\boldsymbol{B}\cdot\boldsymbol{u}=0$, so that $\boldsymbol{K}\cdot\boldsymbol{B}=\boldsymbol{k}\cdot\boldsymbol{B}$. 
We then defined $\boldsymbol{\bar{K}}$ and $\boldsymbol{\bar{B}}$ as the unit vectors along $\boldsymbol{K}$ and $\boldsymbol{B}$ respectively. As $\boldsymbol{k}\cdot\boldsymbol{u}=-\omega$, with $\omega$ the positive pulsation of the photon as measured
by the emitter, $\boldsymbol{\bar{K}}=\boldsymbol{K}/\omega$. As for $\boldsymbol{\bar{B}}$, considering that $\boldsymbol{u}=u^t(\boldsymbol{\partial}_t+\chi\boldsymbol{\partial}_r+\Omega\boldsymbol{\partial}_\varphi)$, with $\chi=u^r/u^t$ and $\Omega=u^\varphi/u^t$, via equation \eqref{eq:B7}, it is straightforward to get\footnote{We note that for a pure Keplerian velocity, $\chi=0$ and so $\boldsymbol{\bar{B}}_\text{Radial}=\mathbf{e}_r$.}
\begin{gather*}
	\begin{aligned}
		&\boldsymbol{\bar{B}}_\text{Vertical}=-\mathbf{e}_\theta\\
		&\boldsymbol{\bar{B}}_\text{Radial}=\frac{1}{\sqrt{-(g_{tt}+\chi^2g_{rr})}}(\,\chi\sqrt{g_{rr}}\,\mathbf{e}_t+\sqrt{-g_{tt}}\,\mathbf{e}_r)\:.
	\end{aligned}\tag{C1} \label{eq:C1}
\end{gather*}
It then follows that 
\begin{gather*}
	\begin{aligned}
		&(\boldsymbol{\bar{K}}\cdot\boldsymbol{\bar{B}})_\text{Vertical}=-\frac{k^\theta}{\omega}\sqrt{g_{\theta\theta}}\\
		&(\boldsymbol{\bar{K}}\cdot\boldsymbol{\bar{B}})_\text{Radial}=\left(\,\frac{k^r}{\omega}-\chi\,\frac{k^t}{\omega} \right)\frac{\sqrt{-g_{tt}g_{rr}}}{\sqrt{-(g_{tt}+\chi^2g_{rr})}}\:.
	\end{aligned}\tag{C2} \label{eq:C2}
\end{gather*}
The synchrotron pitch angle is defined from these quantities as
\begin{align*}
	\theta_\mathrm{B}&=\arccos(\boldsymbol{\bar{K}}\cdot\boldsymbol{\bar{B}})
	\:.\tag{C3} \label{eq:C3}
\end{align*}

At very large radial distance on the disc, the emitter can be approximated as static, so that $\chi=0$ and $\omega=-\boldsymbol{k}\cdot\boldsymbol{u}=-g_{tt}k^{t}$. Also, since far from the black hole the space-time is described by the Minkowski metric, with $g_{tt}=-1$ and $g_{rr}=1$, we got $\omega=k^t$ and $k^t$ could be related to the photon energy defined in equation \eqref{eq:1}: $k^t=g^{tt}k_t=-E/(g_{tt}\hbar)=E/\hbar$. Using all these remarks, equations \eqref{eq:C2} drastically simplify in the asymptotic limit $r\to\infty$: $(\boldsymbol{\bar{K}}\cdot\boldsymbol{\bar{B}})^\infty_\text{Vertical}=-\hbar rk^\theta/E$ and $(\boldsymbol{\bar{K}}\cdot\boldsymbol{\bar{B}})^\infty_\text{Radial}=\hbar k^r/E$. Then, noticing that for a face-on observer the angular momentum, $L$, of equation \eqref{eq:1} is zero for all photons, we could retrieve the expressions of $k^\theta$ and $k^r$ by taking the vanishing spin parameter, $a$, in the Kerr equations of motion \citep{gralla2020null}. We finally got $(\boldsymbol{\bar{K}}\cdot\boldsymbol{\bar{B}})^\infty_\text{Vertical}=\mp_{\theta} \sqrt{\bar{Q}}/r$, with the minus sign corresponding to a photon emitted downwards in $\theta$, and $(\boldsymbol{\bar{K}}\cdot\boldsymbol{\bar{B}})^\infty_\text{Radial}=\pm_r\sqrt{1-\bar{Q}(r^2-2Mr)/r^4}$, with the minus sign corresponding to a photon emitted towards the black hole radially. For an on-axis observer, the reduced Carter constant, $\bar{Q}$, is directly related to the polar radius, $b$, on the screen \citep{gralla2020lensing} via $\bar{Q}=b^2$ and the approximation \eqref{eq:3}, $b\approx r+M$, then yields $\bar{Q}\approx r^2+2Mr+M^2$. Thus, for $n=0$ photons, $(\boldsymbol{\bar{K}}\cdot\boldsymbol{\bar{B}})^{\infty,\,n=0}_\text{Vertical}=-1\Rightarrow (\sin\theta_\mathrm{B})^{\infty,\,n=0}_\text{Vertical}=0$ and $(\boldsymbol{\bar{K}}\cdot\boldsymbol{\bar{B}})^{\infty,\,n=0}_\text{Radial}=0\Rightarrow (\sin\theta_\mathrm{B})^{\infty,\,n=0}_\text{Radial}=1$. For $n=1$ photons, it is clear from Figure \hyperref[fig:thetaB]{C.3} that the radial growth of $\bar{Q}$, unchanged along one geodesic, is much slower than $r^2$ with respect to the second impact radius. As a consequence, $(\boldsymbol{\bar{K}}\cdot\boldsymbol{\bar{B}})^{\infty,\,n=1}_\text{Vertical}=0\Rightarrow (\sin\theta_\mathrm{B})^{\infty,\,n=1}_\text{Vertical}=1$ and $(\boldsymbol{\bar{K}}\cdot\boldsymbol{\bar{B}})^{\infty,\,n=1}_\text{Radial}=-1\Rightarrow (\sin\theta_\mathrm{B})^{\infty,\,n=1}_\text{Radial}=0$, where the ingoing sign can be understood from Figure \hyperref[fig:thetaB]{C.3}.\\
\indent When radially approaching the event horizon, no trivial\footnote{Except for the conditions $u^\theta=0$ and $L=p_\varphi=g_{\varphi\varphi}p^\varphi=0$, translating an equatorial disc motion and a face-on view respectively.} simplifications apply to $\omega=-g_{tt}k^tu^t-g_{rr}k^ru^r$. However, again from the geodesic equations \citep{gralla2020null}, the asymptotic behaviour of $w$ is $w^{r_\mathrm{H}}=[u^t-u^r r/(r-2M)]E/\hbar$. Then, using the fact that $u^t>0$ for future-directed particles and $u^r$ is negative and non-vanishing below the radius of the ISCO, it is straightforward to conclude that $\omega\to+\infty$ at the horizon. As $k^\theta$ is finite at the horizon, for a vertical magnetic field, we find $(\boldsymbol{\bar{K}}\cdot\boldsymbol{\bar{B}})^{r_\mathrm{H},\,n=0,1}_\text{Vertical}=0 \Rightarrow (\sin\theta_\mathrm{B})^{r_\mathrm{H},\,n=0,1}_\text{Vertical}=1$. In the case of a radial magnetic field, we show that the asymptotic limit of equation \eqref{eq:C2} is $(\boldsymbol{\bar{K}}\cdot\boldsymbol{\bar{B}})^{r_\mathrm{H},\,n=0,1}_\text{Radial}=\varepsilon_\mathrm{sub}r\left(\varepsilon_\mathrm{sub}^2r^2+C^2\omega_\varphi^2\ell_\mathrm{sub}^2\right)^{-1/2}$ where we have introduced $C\coloneqq-\omega_r\varepsilon_\mathrm{sub}-(1-\omega_r)$ and employed the quantities $\varepsilon_\mathrm{sub}, \ell_\mathrm{sub}, \omega_r$, and $\omega_\varphi$ presented in Paragraph \hyperref[sec:mixed]{4.1.3}. For a sub-Keplerian motion, for which $\omega_r=\omega_\varphi=1$, this quantity reduces to $r\left(r^2+\ell_\mathrm{sub}^2\right)^{-1/2}=\frac{1}{2}[(4-\xi^2)/(1+2\xi^2)]^{1/2}$ which is equal to 0.5 for Keplerian velocity profile with $\xi=1$ (see Paragraph \hyperref[sec:mixed]{4.1.3}). For a radial infall, the previous scalar product becomes 1 as $\omega_r=\omega_\varphi=0$. To summarise, given the values of Table \hyperref[tab:astro]{2}, $(\sin\theta_\mathrm{B})^{r_\mathrm{H},\,n=0,1}_\text{Radial}\simeq0.87$, 0 or 0.67 for a Keplerian, radial or mixed velocity profile respectively.\\
\indent Finally, we considered the intermediate extrema of the $\sin\theta_\mathrm{B}$ function. Generally, an extremum of $\sin\theta_\mathrm{B}$ is found when $\cos\theta_\mathrm{B}=0\Leftrightarrow \boldsymbol{\bar{K}}\cdot\boldsymbol{\bar{B}}=0$ or when $\boldsymbol{\bar{K}}\cdot\boldsymbol{\bar{B}}\neq\pm1$ and $d(\boldsymbol{\bar{K}}\cdot\boldsymbol{\bar{B}})/dr=0$. The first condition, corresponding to a maximum as $\sin\theta_\mathrm{B}=1$, is never verified for a vertical magnetic field, as $\omega$ is finite and $k^\theta\neq0$ for non-extremal values of $r$, and it is satisfied, in the case of a radial magnetic field for $k^r/k^t=\chi=u^r/u^t$. We numerically show that, for all the velocity profiles investigated in this paper, this relation is never verified for the $n=0$ photons and only for a single value of $r$, depending on the disc velocity, in the case of $n=1$ photons. In particular, $r\simeq3.6$, 5.6 or $4.4M$ in the case, respectively, of a Keplerian, radial or mixed velocity with the parameters of Table \hyperref[tab:astro]{2}. In the case of a vertical magnetic field, using Leibniz rule and the geodesic equations \citep{gralla2020null}, the second condition can be rewritten as $r=-\omega/(d\omega/dr)$. We note that this condition is never verified for $n=0$ photons, whereas it leads to a single\footnote{As $(\sin\theta_\mathrm{B})^{n=1}_\text{Vertical}$ is not a constant function and it tends to 1 both at infinity and at the horizon, we already knew that it possesses at least one minimum; here, we claim the unicity of this extremum.} minimum position $r>2M$ for $n=1$ photons: $r\simeq3.9$, 5.3 or $4.6M$ for a Keplerian, radial or mixed velocity, respectively. For a radial magnetic field, no trivial simplification of the second condition can be easily found, but we notice that an additional minimum is present for the Keplerian and mixed velocity profiles both for the $n=0$ and $n=1$ photons at, respectively, $r\simeq2.6M$ and $r\simeq2.7M$ or $r\simeq2.2M$ and $r\simeq2.3M$.\\
\indent Here, we summarise our findings and their implications for the position of the peaks of intensity. For weakly lensed photons and a radial magnetic field, $\sin\theta_\mathrm{B}$ increases from an intermediate value to 1 with increasing distance from the central object, after reaching a minimum in the case of a Keplerian or mixed velocity, while it decreases from 1 to 0 in the presence of a vertical magnetic field. As a consequence, $j_\nu$ given by equation \eqref{eq:21}, at a fixed $r$, is smaller, in both cases, than when taking $\theta_\mathrm{B}=90^\circ$, but more importantly, the emission profile dies faster for a radial magnetic field and even more so for a vertical one. This implies that, when incorporating redshift effects, the $n=0$ peak positions will be reached at further radii for the $\theta_\mathrm{B}=90^\circ$ profiles than for the ones where $\theta_\mathrm{B}$ is computed numerically. Specifically, the maxima occur at smaller radii in the case of a radial magnetic field and even smaller radii for a vertical field. Part of this reasoning is confirmed by comparing the left columns of  Figures \hyperref[fig:peaksnomag]{7} and \hyperref[fig:peaksnomag]{8}.\\
\indent Concerning the first-lensed photons, whose geodesics are illustrated in light blue in Figure \hyperref[fig:thetaB]{C.3}, a global extremum in $\sin\theta_\mathrm{B}$ is present in all scenarios, being a minimum for a vertical magnetic field and a maximum for a radial one (see right inset of Figure \hyperref[fig:thetaB]{C.3} for a mixed velocity profile). Then, in the radial-$B$ case, inner emission below the maximum is weakened even before the redshift knock down detailed in Section \hyperref[sec:analysis]{6}. This explains the behaviour remarked in Figure \hyperref[fig:RadialB]{C.2}: $n=1$ peak positions appear further away for a radial magnetic field. Similarly, as the inner emission for a vertical magnetic field is boosted, $n=1$ peaks appear at smaller radii than the ones of profiles with constant $\theta_\mathrm{B}=90^\circ$ (see right columns of  Figures \hyperref[fig:peaksnomag]{7} and \hyperref[fig:peaksnomag]{8}).\\
\indent The integration of $\theta_\mathrm{B}$ in the thermal synchrotron emission profile is then an essential ingredient to take into account when interpreting observed images and its imprints on lensed observables could provide some hints on the magnetic field structure. 

\subsection{\texorpdfstring{$r_\mathrm{inner}$}{rinner}}
\label{app:rinner}

As already pointed out in Section \hyperref[sec:disentangle]{7.1}, the key ingredient to break the degeneracy with the peak positions of the first photon ring is the different size of the critical curve in various space-times. The question now is to know how robust this result is when varying the position of the inner radius of the disc. For example, by taking a RZ space-time with $\epsilon\,|_{\,a_1=0}=0.1$ and a disc with an inner radius $r_\mathrm{inner}=2M$ instead of $1.81M$, coinciding with the horizon of a Schwarzschild black hole, the violet band of the bottom right panel of Figure \hyperref[fig:peaksmag]{8} is shifted upwards by at most 0.05 µas, preserving the conclusions about possible observable disentanglements. However, already for $r_\mathrm{inner}=3M$, the violet band is just 0.1 µas below the pink one, meaning that the impacts of $a_1$ and $\epsilon$ become comparable.\\
\indent The position of the inner radius of the disc is even more crucial for the analysis of the primary image. While variations in the critical curve do not impact the $n=0$ peak positions, they are primarily governed by the inner radius, where the considered emission profiles reach their maximum intensity. For instance, in the case, again, of  a RZ space-time with $\epsilon\,|_{\,a_1=0}=0.1$ and a disc with $r_\mathrm{inner}=2M$, the violet band of the bottom left panel of Figure \hyperref[fig:peaksmag]{8} gets closer to the pink one with a shift of about 0.25 µas, putting the deviations on $a_1$ and $\epsilon$ on a more similar footing. For $r_\mathrm{inner}=3M$, the upper part of the Schwarzschild green band is encompassed.\\
\indent Hence, the position of the inner edge of the accretion disc could be an important source of uncertainty, especially for the interpretation of the direct image. Nevertheless, for MAD discs, the presence of dynamically important near-horizon magnetic fields ensures that the plasma moves slowly all the way down to the black hole horizon \citep{narayan2003magnetically,EHTVIII}, thus taking $r_\mathrm{inner}=r_\mathrm{H}$ is a fair choice.

\subsection{Inclination}
\label{app:inclination}
\begin{figure}[ht!]
	\label{fig:incli}
	\centering
	\includegraphics[width=\hsize]{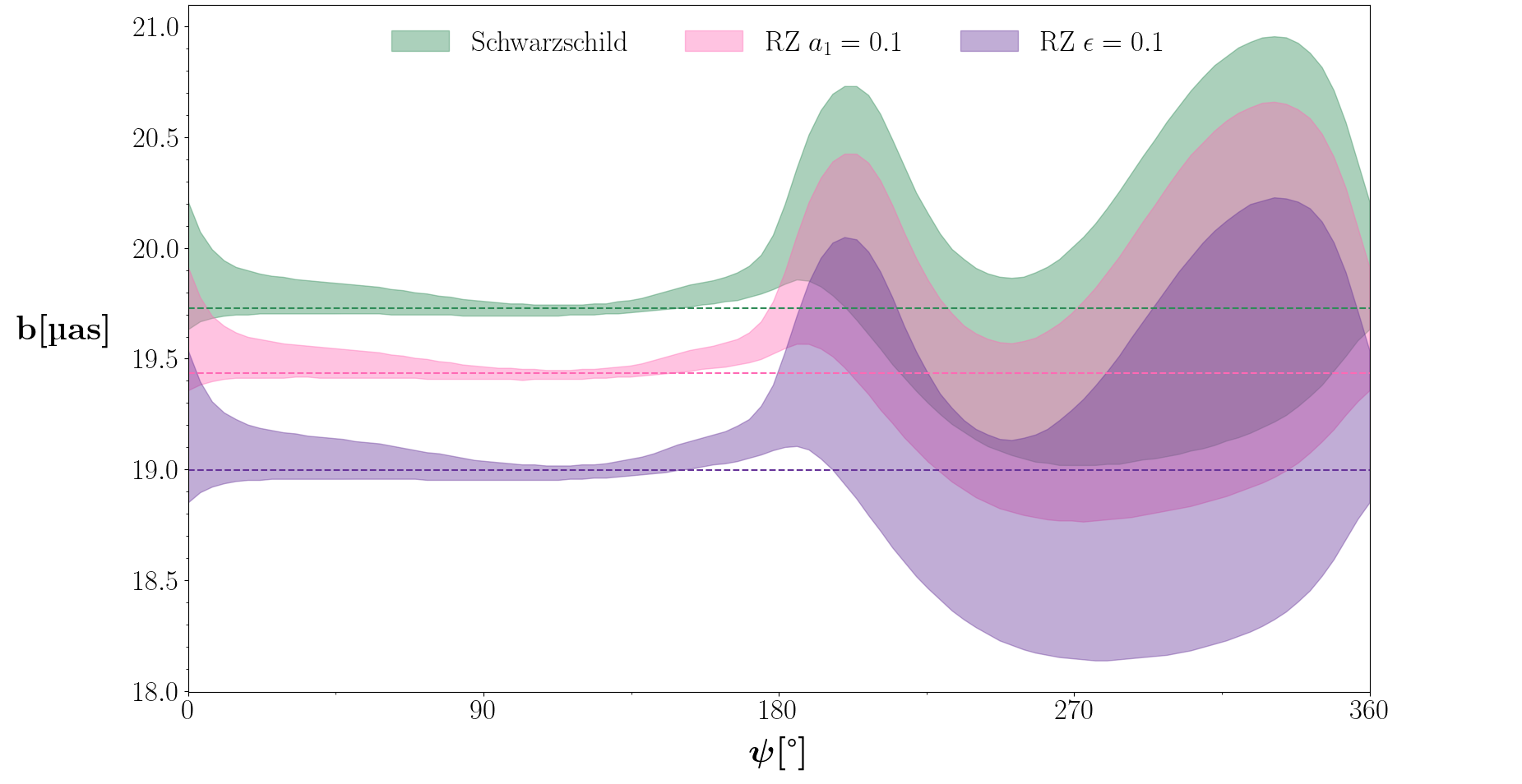}
	\caption{Positions of the $n=1$ peaks, observed at $i=100^\circ$, for all possible emission configurations in the case of a Schwarzschild black hole (green), a RZ black hole with $a_1|_{\,\epsilon=0}=0.1$ (pink), and a RZ black hole with $\epsilon\,|_{\,a_1=0}=0.1$ (violet). Critical curves are dashed lines.}
\end{figure}

We analysed the contribution of a strong inclination, namely $i=100^\circ$, to the study of the degeneracy on the peaks of the 1D intensity profiles. Although this viewing angle is not applicable to M87* or Sgr A* \citep{akiyama2022first}, the investigations of this paragraph could be useful when, with the future interferometric BHEX observations, a broader number of supermassive black holes will be within reach \citep{BHEX}.
To properly account for the deformation due to projection effects, extensively discussed in Appendix \hyperref[app:lensing_bands]{A}, the edges of the first lensing band were ray-traced on 100 polar angles on the screen instead of 14 as in Section \hyperref[sec:simu]{5}; on the other hand, the radial resolution was unchanged.\\
\indent It is interesting to notice that, contrary to the low-inclination framework treated in the main text, here there are some privileged polar directions on the screen to look for metric deviations. In Figure \hyperref[fig:incli]{C.4}, the three bands are clearly separated on the left side, while they are widely superposed on the right.\\
\indent As it is quite evident from Figure \hyperref[fig:incli]{C.4}, the fractional asymmetry is higher than for $i=163^\circ$, reaching almost 40\% for the primary image and 3\% for the first photon ring. Those values are nevertheless degenerate between the various space-times as well as those of the size ratios.

\subsection{\texorpdfstring{$a_0$}{a0} and \texorpdfstring{$b_0$}{b0}}

The zeroth-order parameters of the RZ metric of Section \hyperref[sec:metric]{3} are directly related to the PPN parameters $\beta$ and $\gamma$ via the following equations, where $\epsilon\coloneqq2M/r_{\mathrm{H}}-1$ \citep{rezzolla2014new}:
\vspace{-0.01cm}
\begin{align*}
	a_0=\frac{(\beta-\gamma)(1+\epsilon)^2}{2} \quad\text{and}\quad b_0=\frac{(\gamma-1)(1+\epsilon)}{2}\:.\tag{C5} \label{eq:C5}
\end{align*}
The observational constraints on the PPN parameters from Solar System measurements \citep{will2014confrontation}
\begin{align*}
	|\beta-1|\leq2.3\times10^{-4}
	\quad\text{and}\quad
	|\gamma-1|\leq2.3\times10^{-5}\:,\tag{C6} \label{eq:C6} 
\end{align*}
force then $a_0$ and $b_0$ to be small: $|a_0|\leq10^{-4}$ and 
$|b_0|\leq10^{-4}$. Moreover, the values of $\beta$ and $\gamma$, and therefore those of $a_0$ and $b_0$, are corroborated by stellar astrometry results in the galactic centre \citep{abuter2020detection,dayem2024improving}. Since the galactic centre also hosts a supermassive black hole, assuming that the PPN constraints hold in the surroundings of M87* too, as we did in the paper, is a reasonable choice.  

\begin{figure}[ht!]
	\label{fig:a0b0b1}
	\centering
	\includegraphics[width=\hsize]{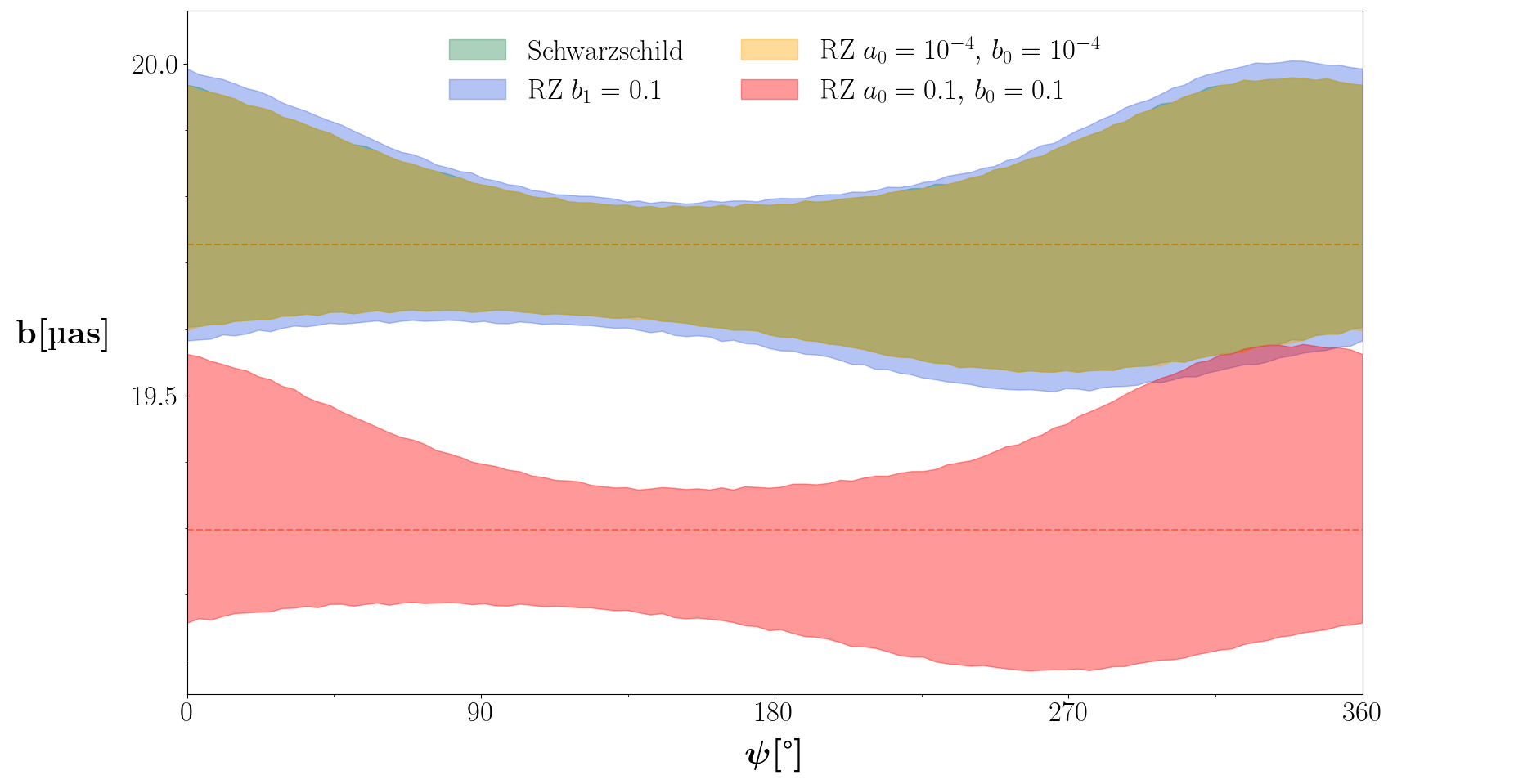}
	\caption{Positions of the $n=1$ peaks for all the possible emission configurations in the case of a Schwarzschild black hole (green), a RZ black hole with non-vanishing parameter $b_1=0.1$ (blue), and a RZ black hole with non-vanishing parameters $a_0,\,b_0=10^{-4}$ (yellow) or $a_0,\,b_0=0.1$ (red). Dashed lines represent critical curves: the green and blue ones are superposed while the yellow one is slightly above them.}
\end{figure}

If PPN constraints are satisfied, fixing $a_0$ and $b_0$ to zero, as we chose to do in this article, is legit and useful to analyse the impact of more influential parameters. In fact, as can be seen from the orange bands of Figure \hyperref[fig:a0b0b1]{C.5}, comprising the positions of the intensity peaks for all the emission profiles of Table \hyperref[tab:astro]{2}, a space-time defined by the maximum allowed values of $a_0$ and $b_0$ from Solar System observations, that is, $a_0=b_0=10^{-4}$ with all other parameters vanishing, produces indiscernible differences from the Schwarzschild green case.\\
\indent If PPN constraints are not satisfied around M87*, which is possible in theories of gravity without a Birkhoff-like uniqueness theorem, the $a_0$ parameter is a relevant deviation to consider as it leads to observables comparable to those produced by $a_1$ (see Figure \hyperref[fig:a0b0b1]{C.5}). 

\subsection{\texorpdfstring{$b_1$}{b1}}
\label{app:b1}
We repeated the same study on the peaks positions for small deviations in the parameter $b_1$. Particularly, we chose $b_1=0.1$, similarly to what was done for $\epsilon$ or $a_1$, and we fixed all the other metric parameters to their Schwarzschild value. The results of this space-time configuration are contained in the blue band of Figure \hyperref[fig:a0b0b1]{C.5}, which does not significantly stand out from the green Schwarzschild band. In other words, small geometrical deviations in $b_1$ are degenerate with the astrophysical uncertainty both for the primary image and for the first-lensed photon ring, because $b_i$ parameters, with $i\in\mathbb{N}$, do not affect any of the space-time properties related to the key image features, as explained in Section \hyperref[sec:analysis]{6}. Hence, this justifies our decision to ignore its effects in the main exploration of this article. We note that even for the extreme value $b_1=1$, the degeneracy is never broken.

\subsection{\texorpdfstring{$a_1$}{a1} and \texorpdfstring{$\epsilon$}{e}}
\label{app:higher}

\begin{figure*}
	\label{fig:widths}
	\resizebox{\hsize}{!}         {\includegraphics{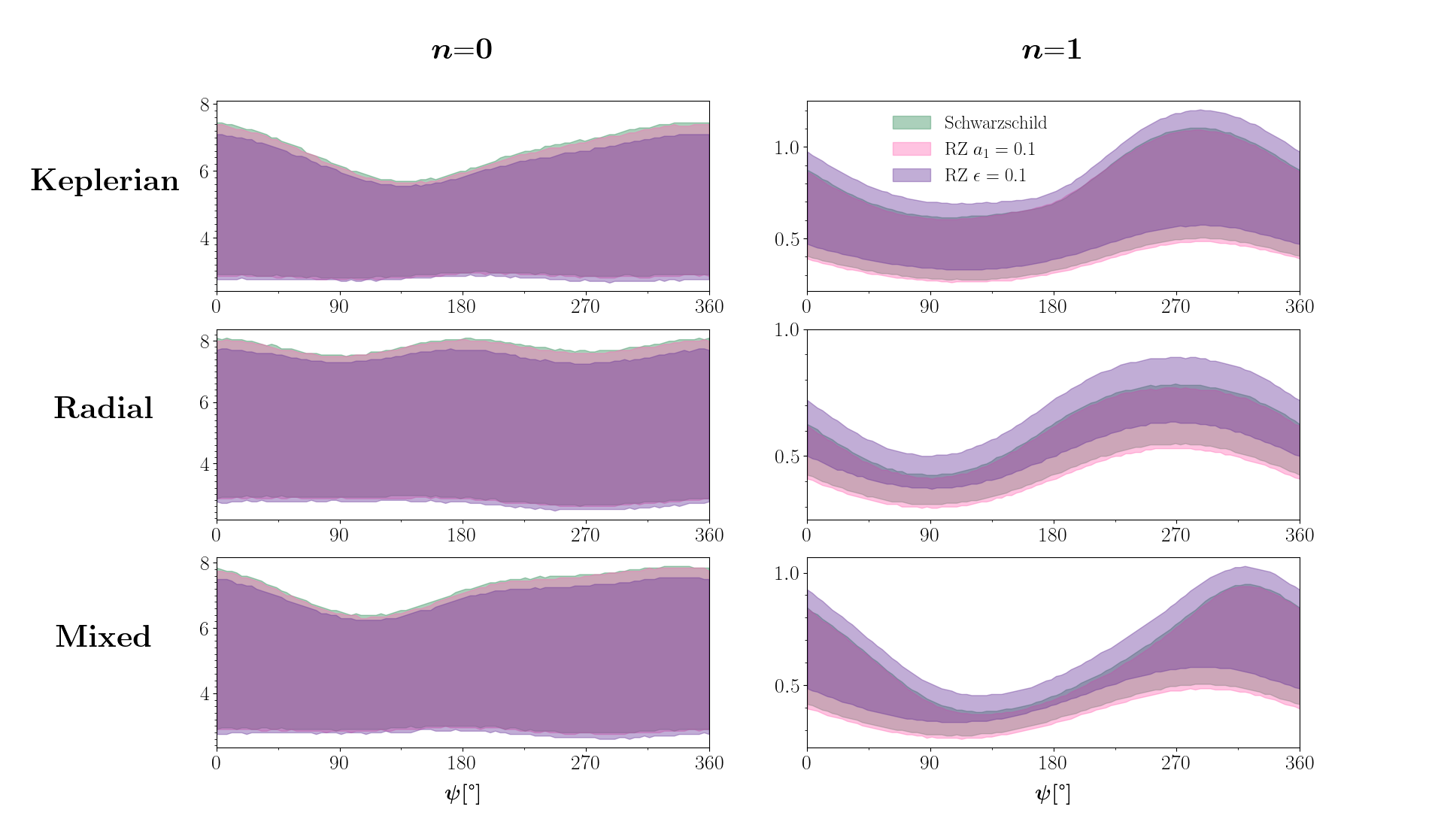}}
	\caption{Widths of the $n=0$ (left panels) and $n=1$ (right panels) images for all the possible emission configurations in the case of a Schwarzschild black hole (green), a RZ black hole with $a_1|_{\,\epsilon=0}=0.1$ (pink), and a RZ black hole with $\epsilon\,|_{\,a_1=0}=0.1$ (violet). The velocity of the disc is taken to be Keplerian for the upper panels, radial for the central panels, and mixed for the lower panels.}
\end{figure*}

\begin{figure*}
	\label{fig:higher}
	\resizebox{\hsize}{!}         {\includegraphics{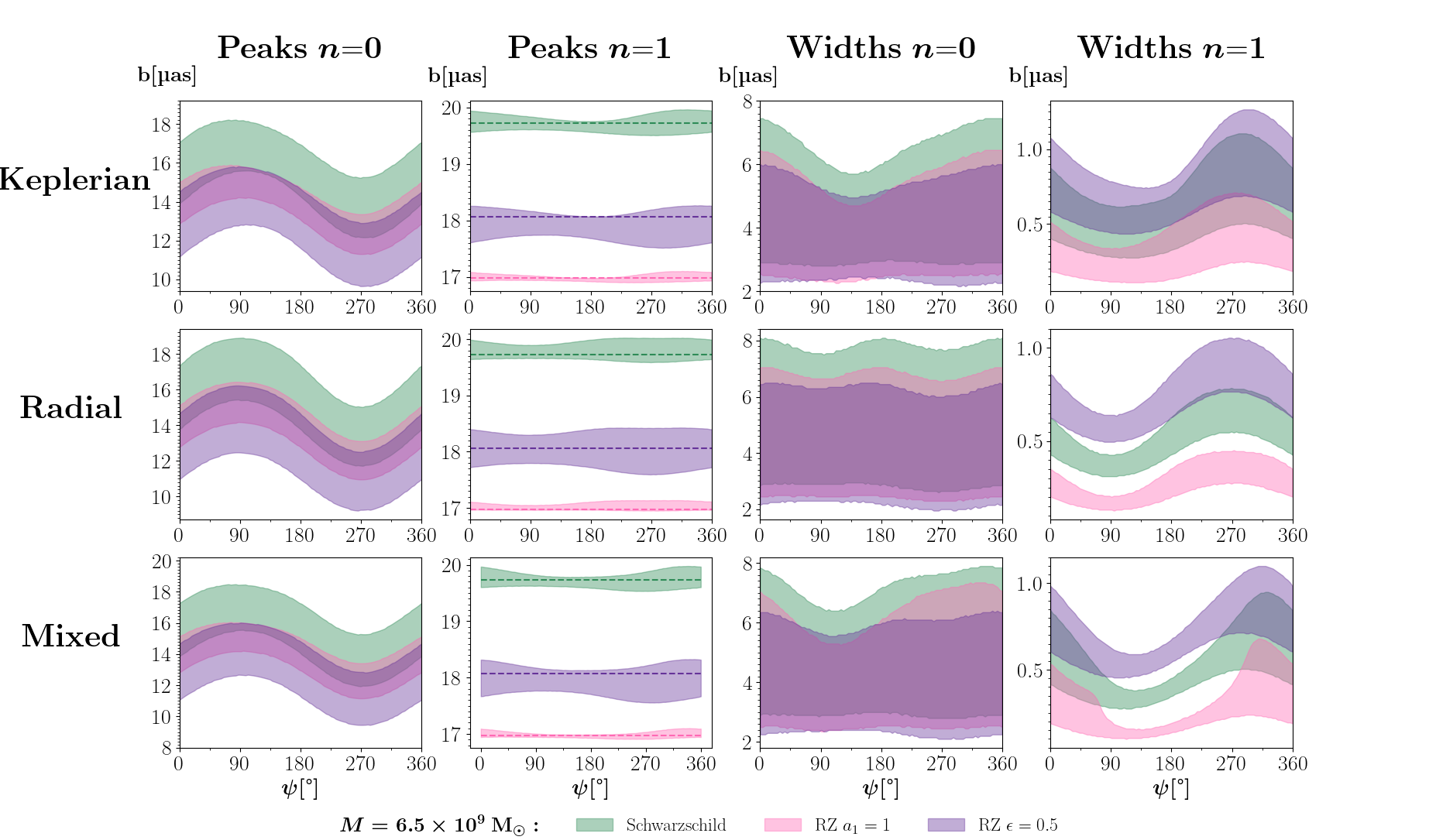}}
	\caption{Positions of peaks and widths of the $n=0$ (odd panels) and $n=1$ (even panels) profiles for all the possible emission configurations in the case of a Schwarzschild black hole (green), a RZ black hole with $a_1|_{\,\epsilon=0}=1.0$ (pink), and a RZ black hole with $\epsilon\,|_{\,a_1=0}=0.5$ (violet). The velocity of the disc is taken to be Keplerian for first row, radial for second row, and mixed for the third row. Critical curves are represented as dashed lines.}
\end{figure*}

For small geometric deviations in $\epsilon$ and $a_1$, the width of the first photon ring is not an observable that permits detection of deviations from the Schwarzschild geometry, as can be seen in Figure \hyperref[fig:widths]{C.6}. However, for higher values of these same parameters, the width can become a complementary probe to the peak positions, although the latter remain a more robust probe. These statements can be visualised in Figures \hyperref[fig:widths]{C.6} and \hyperref[fig:higher]{C.7} where we represented the two extreme positive values allowed by conditions \eqref{eq:7}, namely $a_1|_{\,\epsilon=0}=1$ and $\epsilon\,|_{\,a_1=0}=0.5$. As already discussed in the main text, larger deviations require less stringent $M/D$ precisions to be observed. We remark also that these extreme deviations lead to more striking differences in the extent, but not the shape, of the bands. This is a feature linked to the size of the corresponding lensing bands, which, despite not being observable, must contain the photon rings entirely \citep{cardenas2024lensing}.\\
\indent Even for extreme RZ parametric deviations, the astrophysical-driven degeneracy on the image asymmetry and images size ratios introduced in Section \hyperref[sec:relative]{8.2} is never broken. Thus, relative measurements can never be used to distinguish a Schwarzschild black hole from a RZ space-time.

\end{appendix}

\end{document}